\documentclass{article}

\pdfoutput=1
\usepackage{arxiv}
\usepackage[utf8]{inputenc} 
\usepackage[T1]{fontenc}    
\usepackage{hyperref}       
\usepackage{url}            
\usepackage{booktabs}       
\usepackage{amsfonts}       
\usepackage{nicefrac}       
\usepackage{microtype}      
\usepackage{lipsum}
\RequirePackage{amsmath,amsfonts,amssymb}
\RequirePackage{graphicx,xcolor}
\usepackage[linesnumbered, boxed]{algorithm2e}

\newcommand{\N}{\mathbb{N}}
\newcommand{\Z}{\mathbb{Z}}
\newcommand{\Pro}{\mathbb{P}}
\newcommand{\E}{\mathbb{E}}

\title{\textbf{A}rtificial \textbf{B}enchmark for \textbf{C}ommunity \textbf{D}etection (\textbf{ABCD})\\
Fast Random Graph Model with Community Structure}

\author{
Bogumi\l{} Kami\'nski\thanks{Decision Analysis and Support Unit, SGH Warsaw School of Economics, Warsaw, Poland; e-mail: \texttt{bogumil.kaminski@sgh.waw.pl}}
\And
Pawe\l{}~Pra\l{}at\thanks{Department of Mathematics, Ryerson University, Toronto, ON, Canada; e-mail: \texttt{pralat@ryerson.ca}}
\And
Fran\c{c}ois Th\'eberge\thanks{Tutte Institute for Mathematics and Computing, Ottawa, ON, Canada; e-mail: \texttt{theberge@ieee.org}}
}

\begin{document}

\maketitle

\begin{abstract}
Most of the current complex networks that are of interest to practitioners possess a certain community structure that plays an important role in understanding the properties of these networks. For instance, a closely connected social communities exhibit faster rate of transmission of information in comparison to loosely connected communities. Moreover, many machine learning algorithms and tools that are developed for complex networks try to take advantage of the existence of communities to improve their performance or speed. As a result, there are many competing algorithms for detecting communities in large networks.

Unfortunately, these algorithms are often quite sensitive and so they cannot be fine-tuned for a given, but a constantly changing, real-world network at hand. It is therefore important to test these algorithms for various scenarios that can only be done using synthetic graphs that have built-in community structure, power-law degree distribution, and other typical properties observed in complex networks.

The standard and extensively used method for generating artificial networks is the \textbf{LFR} graph generator. Unfortunately, this model has some scalability limitations and it is challenging to analyze it theoretically. Finally, the mixing parameter $\mu$, the main parameter of the model guiding the strength of the communities, has a non-obvious interpretation and so can lead to unnaturally-defined networks.

In this paper, we provide an alternative random graph model with community structure and power-law distribution for both degrees and community sizes, the \textbf{A}rtificial \textbf{B}enchmark for \textbf{C}ommunity \textbf{D}etection (\textbf{ABCD} graph). We show that the new model solves the three issues identified above and more. In particular, we test the speed of our algorithm and do a number of experiments comparing basic properties of both \textbf{ABCD} and \textbf{LFR}. The conclusion is that these models produce comparable graphs but \textbf{ABCD} is fast, simple, and can be easily tuned to allow the user to make a smooth transition between the two extremes: pure (independent) communities and random graph with no community structure.
\end{abstract}

\section{Introduction}

An important property of complex networks is their community structure, that is, the organization of vertices in clusters, with many edges joining vertices of the same cluster and comparatively few edges joining vertices of different clusters~\cite{Paper1, Paper2}. In social networks, communities may represent groups by interest (practical applications include collaborative tagging), in citation networks they correspond to related papers, similarly in the web communities are formed by pages on related topics, etc. Being able to identify communities in a network helps to exploit it more effectively. For example, clusters in citation graphs may help to find similar scientific papers, discovering users with similar interests is important for targeted advertisement, clustering can also be used for network compression and visualization. Finally, many machine learning algorithms and tools use clustering as a unsupervised pre-processing step and then try to take advantage of the community structure to improve their performance or speed.

The goal of community detection algorithms is to partition the vertex set of a graph into subsets of vertices called communities such that there are more edges present within communities in comparison to the global density of the graph. The key ingredient for many clustering algorithms is modularity. Modularity for graphs was introduced by Newman and Girvan~\cite{Modularity_Newman} and it is based on the comparison between the actual density of edges inside a community and the density one would expect to have if the vertices of the graph were attached at random regardless of community structure, while respecting the vertices’ degree on average. There are many variants allowing, in particular, overlapping or hierarchical communities. Moreover, it is also possible to generalize modularity for hypergraphs~\cite{Modularity_Pralat}.

Unfortunately, detecting communities in networks is a challenging problem. Many algorithms and methods have been developed over the last few years---see, for example, \cite{dao_bothorel_lenca} for a recent review. It is important to point out that these algorithms are often quite sensitive and so they cannot be fine-tuned for a given family of networks we want these algorithms to work on. Some algorithms perform well on networks with strong communities but perform poorly on graphs with weak communities. The degree distribution and other properties of networks may also drastically affect the performance of these algorithms in terms of accuracy or computational complexity. Because of that it is important to test these algorithms for various scenarios that can only be done using synthetic graphs that have built-in community structure, power-law degree distribution, and other typical properties observed in complex networks.

In order to compare algorithms, one can use some quality measure, for example, the above mentioned  modularity~\cite{Modularity_Newman}. Indeed, modularity is not only a global criterion to define communities and a way to measure the presence of community structure in a network but, at the same time, it is often used as a quality function of community detection algorithms. However, it is not a fair benchmark, especially for comparing algorithms (such as Louvain and Ensemble Clustering) that find communities by trying to optimize the very same modularity function! In order to evaluate algorithms in a fair and rigorous way, one should compare algorithm solutions to a synthetic network with an engineered ground truth.

The standard and extensively used method for generating artificial networks is the \textbf{LFR} (Lancichinetti, Fortunato, Radicchi) graph generator~\cite{LFR, LFR2}. This algorithm generates benchmark networks (that is, artificial networks that resemble real-world networks) with communities. The main advantage of this benchmark over other methods is that it allows for the heterogeneity in the distributions of both vertex degrees and of community sizes. As a result, in the past decade, the \textbf{LFR} benchmark has become a standard benchmark for experimental studies, both for disjoint and for overlapping communities~\cite{Emmons}. Some other benchmarks are discussed in the next section.

In order to generate a random graph following a given, previously computed, degree sequence, the \textbf{LFR} benchmark uses the fixed degree sequence model (also known as edge switching Markov chain algorithm) to obtain the desired community structure once the stationary distribution is reached. Unfortunately, the convergence process can be slow and so this model has some scalability limitations. Despite the need for experiments on large networks, the standard \textbf{LFR} implementation\footnote{\texttt{https://github.com/eXascaleInfolab/LFR-Benchmark\_UndirWeightOvp/}}
can only be used to generate medium size networks. Moreover, due to its complexity and the fact that many implementations stop the switching before the stationary distribution is reached, it is challenging to analyze the model theoretically. Finally, the mixing parameter $\mu$, the main parameter of the model guiding the strength of the communities, has a non-obvious interpretation and so can lead to unnaturally-defined networks. We discuss these issues at length in the next section.

\medskip

In this paper, we provide an alternative random graph model with community structure and power-law distribution for both degrees and community sizes, the \textbf{A}rtificial \textbf{B}enchmark for \textbf{C}ommunity \textbf{D}etection (\textbf{ABCD} graph). We show that the new model solves the three issues identified above. In particular, we test the speed of our algorithm and do a number of experiments comparing basic properties of both \textbf{ABCD} and \textbf{LFR}. The conclusion is that these models produce comparable graphs but \textbf{ABCD} is fast, simple, and can be easily tuned between the two extremes: random graph with no community structure and independent communities.
The Julia package providing an API for generation of \textbf{ABCD} graphs can be accessed at GitHub repository\footnote{\texttt{https://github.com/bkamins/ABCDGraphGenerator.jl/}}. The repository also provides instructions how to set up R and Python to use the package directly from these environments. 
(For reference purposes, if requested, we can also provide a Python implementation of \textbf{ABCD}.)
Finally, a command line interface to the library is provided that allows users to generate \textbf{ABCD} graphs without using an API. 

\medskip

The paper is structured as follows. In the next section, Section~\ref{sec:motivation}, we justify the need for a new benchmark network model.
Section~\ref{sec:model} provides a detailed description of the model.
In order to compare \textbf{ABCD} and \textbf{LFR}, one needs to tune the two mixing parameters to make the corresponding graphs comparable. We explain this process in Section~\ref{sec:comparison}.
Section~\ref{sec:experiments} presents experiments for comparison of the two models (both the speed and basic properties).
Brief conclusion and directions for future work are presented in Section~\ref{sec:conclusion}.
Finally, pseudo-codes of our \textbf{ABCD} generator are presented in the Appendix.

\section{Motivation}\label{sec:motivation}

In the introduction, we already highlighted a few issues with the existing \textbf{LFR} benchmark. In this section, we provide more detailed justification for the need of a new benchmark model.

\subsection{Problem with Scalability}\label{sec:scalability}

In the big data era, there are many massive networks that need to be mined and analyzed. Since such networks cannot be handled in the memory of a single computer, new clustering methods have been introduced for advanced models of computation~\cite{Buzun,Zeng}. These algorithms use hierarchical input representations which implies that the experiments performed on small or medium size benchmark graphs cannot be used to predict the performance on much larger instances~\cite{Emmons}. Unfortunately, many graph clustering benchmark generators currently available are not able to generate the graphs of necessary size~\cite{Bae,Buzun}.

Let us briefly discuss the reason for the leading benchmark not to be scalable. Switching edges in \textbf{LFR} can be viewed as a transition in an irreducible, symmetric, and aperiodic Markov chain. As a result, it  converges to the uniform (stationary) distribution. More importantly, if the maximum degree is not too large compared to the number of edges, then it converges in polynomial time~\cite{Greenhill}. However, despite the fact that these bounds on the mixing time are of theoretical importance, they are not practical even for small graphs. The convergence process is inherently slow and so the model has clear scalability limitations that are known to both academics and practitioners. The fastest published variant of the model that is able to generate large graphs is the external memory algorithm proposed by Hamann \emph{et.\ al.}~\cite{Hamann:2018:IGM:3178547.3230743}.

In order to generate huge graphs, practitioners typically use computationally inexpensive random graph models such as R-MAT~\cite{R-MAT} or the generator of Funke \emph{et.\ al.}~\cite{Funke}. These models might create communities. In fact, it is known that many random graph models naturally create community structure, especially the ones that are geometric in nature~\cite{Clustering_Pralat}. However, they are not suitable for benchmarking purposes as there is no ground truth community structure to compare against. Hence, it is difficult to use them to evaluate clustering algorithms.

Another alternative, based on the scalable Block Two-Level Erd\H{o}s-R\'enyi (\textbf{BTER}) graph generator~\cite{Seshadhri2012}, was recently proposed by Slota \emph{et.\ al.}~\cite{slota_sc2019}. The original model takes into account the desired degree distribution and per-degree clustering coefficient. Since it does not aim to create communities, its edge-generation process is more direct, simpler, and faster than \textbf{LFR}'s. The authors of~\cite{slota_sc2019} try to twist the original model to create a graph that resembles the \textbf{LFR} benchmark.

\medskip

The proposed \textbf{ABCD} model is fast. The experiments we performed imply that \textbf{ABCD} is 40 to 100 times faster than \textbf{LFR} (see Subsection~\ref{sec:speed} for more details). In particular, a graph on 10 million vertices with an average vertex degree of 25 can be generated on a standard desktop computer in several minutes (see Table~\ref{tab:timing} for example timing reports; the \textbf{LFR} algorithm implementation we used would take several hours to generate graphs of similar size). In this paper we concentrate on single threaded \textbf{ABCD} and \textbf{LFR} implementations in order to focus on the theoretical concepts behind \textbf{ABCD}. However, as an outlook for further work it is possible to design a distributed out-of-core implementation of \textbf{ABCD} to generate huge graphs having billions of vertices, similarly like it is done in~\cite{Hamann:2018:IGM:3178547.3230743} for \textbf{LFR}. Indeed, for example, generation of edges within communities can be performed using perfectly parallel approach, as each community is processed independently (see Section~\ref{sec:model} for details).

\subsection{Many Variants and Lack of Theoretical Foundations}\label{sec:theory_motivation}

The most computationally expensive part of the \textbf{LFR} benchmark is edge switching. In each step of this part of the algorithm, two edges are chosen uniformly at random and two of the endpoints are swapped if it removes the loop or parallel edge without introducing new ones. As already mentioned, the process converges to the stationary distribution but it does not converge fast enough for large graphs to be produced. Experimental results on the occurrence of certain motifs in networks~\cite{Paper38}, the average and maximum path length and link load~\cite{Paper19} suggest that $\Theta(m)$ swaps are enough to get close to the desired distribution, where $m$ is the number of edges in the graph. (See also~\cite{Paper44} for further theoretical arguments and experiments.) The constant hidden in the asymptotic notation varies from experiment to experiment and is between 2 and 100. There are also some other heuristic arguments that justify more simplifications of the original algorithm.

There are at least two negative implications of this situation. First of all, there are many variants of this benchmark model and various implementations further modify some steps, either as an attempt to simplify the algorithm or to gain on speed. As a result, one can only create ``\textbf{LFR}-type'' graph and graphs generated by different implementations can have different properties. In fact, even the original formulation of the model leaves some ingredients not rigorously defined. This is certainly not desired for benchmark graphs that should provide a rigorous and fair comparison. Moreover, it creates challenges with reproducing experiments, something that is expected, if not required, when reporting scientific results.

The lack of a simple and clear description of the algorithm has another negative aspect. Despite the fact that the initial work on Erd\H{o}s-R\'enyi model did not aim to realistically model real-world networks, the number of papers on random graphs and their applications to model complex networks is currently exploding. Indeed, in the period after 1999, due to the fact that data sets of real-world networks became abundantly available, their structure has attracted enormous attention in mathematics as well as various applied domains. For example, one of the first articles of Albert and Barab\'asi~\cite{BA} in the field is cited by more than 35,000 times. There are many papers investigating models of complex networks starting with a natural generalization of the Erd\H{o}s-R\'enyi model to a random graph with a given expected degree distribution~\cite{CL2006} to more challenging models such as random hyperbolic graphs~\cite{Krioukov} or spatial preferential attachment graphs~\cite{DBLP:journals/im/AielloBCJP08}. These results are not only interesting from theoretical point of view; they help us better understand the properties and the dynamics of these models by investigating local mechanisms that shape global statistics of the produced network. Despite this fact, there are very few results on theoretical properties of the \textbf{LFR} graph. It is unfortunate, as more research on models with community structure might shed light on how communities are formed and help us design better and faster clustering algorithms.

\medskip

As described in Section~\ref{sec:model}, the proposed \textbf{ABCD} model can be seen as a union of independent random graphs.
As a result, its asymptotic behaviour can be studied with the existing tools in random graph theory. Moreover, \textbf{ABCD} model is natural, relatively easy and straightforward to implement that limits a problem with reproducibility. Nevertheless, for those that look for ``out-of-the-box'' tool, we made it available as GitHub repository with a reference implementation.

\subsection{Communities are Unnaturally-defined for Large Mixing Parameters}\label{sec:ill-defined}

\begin{figure}[ht]
\begin{center}
\includegraphics[width=14cm]{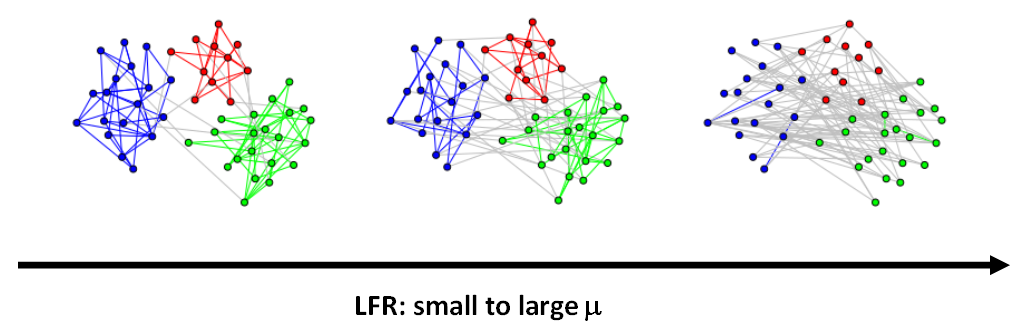}
\vspace{0.5cm}
\includegraphics[width=14cm]{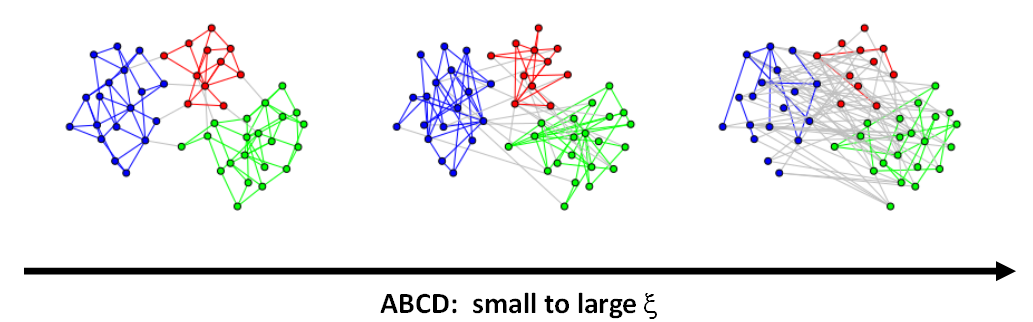}
\end{center}
\caption{Examples of graphs generated by the \textbf{LFR} algorithm (top) and by the \textbf{ABCD} algorithm (bottom). All graphs have the same degree distribution and community sizes. The three \textbf{LFR} graphs correspond to values of the mixing parameter $\mu \in \{0.1, 0.3, 0.95\}$, whereas for the \textbf{ABCD} graphs the plots correspond to $\xi \in \{0.1, 0.3, 0.95\}$.
Edges that fall between vertices in the same community are coloured accordingly.
We see strong communities for the leftmost plots, and noisy yet still coherent communities for the middle plots. The rightmost plots, where $\mu = \xi = 0.95$, illustrate our point regarding one of the main differences between \textbf{LFR} and \textbf{ABCD}. For \textbf{LFR}, in the top right plot, we see almost no edges within each community so the model generates ``anti-communities''. With \textbf{ABCD}, we see a random looking graph, where the number of edges within each ``community'' is proportional to the number of vertices that belong to it, as expected in a random graph.}
\label{fig:mixing}
\end{figure}

One of the parameters of the \textbf{LFR} benchmark is the mixing parameter $\mu \in [0,1]$ which controls the desired ``community tightness''. The goal is to keep the fraction of inter-community edges to be approximately $\mu$. In one of the two extremes, when $\mu = 0$, all edges are within communities. On the other hand, when $\mu=1$, \textbf{LFR} generates pure ``anti-communities'' with no edge present in any of the communities. We believe that this is undesired and leads to unnaturally-defined communities. The threshold value of $\mu$ that produces pure random graph that is community agnostic is ``hidden'' somewhere in the interval $[0,1]$. It is possible to compute this threshold value (see Section~\ref{sec:splitting} where we actually do it) but the formula is quite involved and not widely known. Indeed, many different values are reported in the literature (for example, $\mu=0.7$ is mentioned in~\cite{slota_sc2019}) and so many experiments are performed on unnaturally-defined networks and might lead to false conclusions. The influence of the parameter $\mu$ on the \textbf{LFR} graph is presented in Figure~\ref{fig:mixing} (top).

\medskip

In contrast, the parameter $\xi \in [0,1]$ in the \textbf{ABCD} model (counterpart of $\mu$ in \textbf{LFR} introduced in Section \ref{sec:model}) has natural and important interpretation. As in \textbf{LFR}, if $\xi=0$, then all edges are generated
exclusively within communities. More importantly, $\xi=1$ yields pure random graph in which communities do not affect the process of generating edges. Values of $\xi \in (0,1)$ produce graphs with additional signal coming from communities; the smaller the parameter, the more pronounced the communities are. One can easily move between $\xi$ and $\mu$ (again,  see Section~\ref{sec:splitting} for more details) but there is no risk to create unnaturally-defined benchmark networks with ``anti-communities''. The influence of the parameter $\xi$ on the \textbf{ABCD} graph is presented in Figure~\ref{fig:mixing} (bottom).

\subsection{Densities of Communities}

The \textbf{LFR} model aims to generate a graph in which $(1-\mu)$ fraction of edges adjacent to a given vertex stays within the community of that vertex. This property should hold for all vertices regardless whether this vertex belongs to large or small community. As a result, small communities will become much denser comparing to large ones. It is not clear that this property is desirable, especially in the case of unbalanced community sizes which the model is aiming for. Indeed, it seems to us that larger clusters should capture a proportionally larger fraction of edges---see Subsection~\ref{sec:global_vs_local} for a detailed discussion.

The approach used in \textbf{LFR} (which we call a \textbf{local} variant) seems to be inherited from the definition of the community in the classical book of Barab\'asi~\cite{Barabasi2016}. We challenge it and propose another approach (that we call a \textbf{global} variant) although we do respect this point of view. For those researchers and practitioners who prefer the original approach, we describe two variants of the \textbf{ABCD} model, one for each approach, and both variants are available on GitHub repository.

\section{Proposed Model}\label{sec:model}

\subsection{Parameters of the Model}

We assume that the following parameters are provided as the input for the algorithm (for each input we specify a general approach and a proposed default specification):
\begin{enumerate}
    \item The number of vertices, $n$. \\ (Notation: We label vertices with  numbers from $V=[n]:=\{1, \ldots, n\}$.)
    \item The exact (or expected) degree distribution $\mathbf{w}=(w_1, \ldots, w_n)$. The user can decide if the degree distribution has to follow a given distribution exactly (the configuration model will be used in this case) or only to follow it in expectation (the Chung-Lu model will be used instead). \\
    (Remark: Note that the user does not have to provide vector $\mathbf{w}$ explicitly; it could be generated at random. In particular, it could follow a power law distribution with parameter $\gamma$ and extreme values $w_{\min}$, $w_{\max}$. Alternatively, the average value $\bar{w}$ can be supplied instead of $w_{\min}$, which can then be computed, as it is done in the original \textbf{LFR} model.)
    \item The number of clusters, $k$, and the sequence of cluster sizes $\textbf{s} = (s_1, \ldots, s_k)$ satisfying $\sum_{i=1}^k s_i = n$. \\ (Remark: In particular, $\textbf{s}$ could be a random sequence following the power law distribution with parameter $\beta$ and extremes $s_{\min}$, $s_{\max}$, as it is done in \textbf{LFR}. If not specified, by default \textbf{LFR} sets $s_{\min}$ and $s_{\max}$ to the minimum and, respectively, the maximum degree.) \\ (Notation: We label clusters with numbers from $[k]$. We will use $f(\sigma(i)) \in [k]$ to denote the cluster of vertex $i\in[n]$, see Subsection~\ref{sec:assignment} for a formal definition of this mapping.)
    \item The mixing parameter $\xi$. \\
    (Remark: As already mentioned, at one extreme case when $\xi=0$, all links are within clusters. On the other hand, if $\xi=1$, then communities do not influence distribution of edges. Moreover, to add more flexibility, one may introduce different parameters $\xi$ for each cluster---see Subsection~\ref{sec:global_vs_local} for more on that.)
\end{enumerate}

\subsection{Sampling w and s}
\label{sec:sampling}

At the very beginning, we sample the exact/expected degree distribution $\mathbf{w}$, the number of clusters, and the cluster sizes $\textbf{s}$ (unless they are given as deterministic parameters of the model). The algorithms used to generate them are presented in Appendix. Let us stress the fact that if $\textbf{w}$ and $\textbf{s}$ are sampled, then they are random variables. However, for fair comparison purposes, the same values of $\mathbf{w}$ and $\mathbf{s}$ are used when experiments on \textbf{LFR} and \textbf{ABCD} models are performed in Section \ref{sec:experiments}.

\subsection{Background and Cluster Graphs}

Our model can be viewed as a union of $k+1$ independent random graphs $G_i$ ($i \in [k] \cup \{0\}$)---one for each cluster, and one for the whole graph. As a result, one can view it as a generalization of the \emph{double round exposure method} (also known in the literature as ``\emph{sprinkling}''). We start with the background graph $G_0$ and ``sprinkle'' additional edges within communities that come from graphs $G_i$ ($i \in [k]$); the smaller value of $\xi$, the stronger ties between members of the same cluster are. As these graphs are generated independently, one can alternatively start with the cluster graphs and then ``sprinkle'' the background graph on top of it that can be seen as adding the ``noise''; the larger value of $\xi$, the larger level of noise is.

\medskip

First, we need to split the weight vector $\textbf{w}$ into $\textbf{y}$ and~$\textbf{z}$; $\textbf{z}$ will be responsible for the background graph and $\textbf{y}$ will affect additional edges within communities. The process of splitting the weight is discussed in Section~\ref{sec:splitting}. Then, for a given cluster $i \in [k]$, we restrict ourselves to $V_i \subseteq V =  [n]$, the set of vertices that belong to cluster $i$. We discuss the process of assigning vertices into clusters in Section~\ref{sec:assignment}. Let $\textbf{y}_i$ be the sub-sequence of $\textbf{y}$ restricted to terms corresponding to vertices from $V_i$. Let $G_i = (V_i, E_i)$ be a random graph $\textbf{G}(\textbf{y}_i)$ guided by the sequence $\textbf{y}_i$---the exact model will be specified in Section~\ref{sec:two_models}. Finally, let $G_0=(V,E_0)$ be a random graph $\textbf{G}(\textbf{z})$ guided by the sequence $\textbf{z}$. We call graph $G_0$ the \textbf{background graph} and the remaining graphs $G_i$ (for $i \in [k]$) are called the \textbf{cluster graphs}. The model is defined as the union of these graphs, that is, $\textbf{G} = (V,E)$, where $E = \bigcup_{i=0}^{k} E_i$.

\medskip

Note that $\textbf{G}$ allows loops and multiple edges. Indeed, they can occur both in any of the generated graphs $G_i$ ($i \in [k] \cup \{0\}$) or after taking a union of their edge sets. In general, however, there will not be very many of them, especially for sparse graphs. If one wants to study this random graph theoretically, one option is to work with multi-graphs or condition on $\textbf{G}$ to be simple. From practical point of view, one can still work with multi-graphs or do some minor adjustments to the graph such as rewiring, re-sampling, or simply delete parallel edges. We will come back to this practical issue and provide a specific solution in Section~\ref{sec:two_models}. However, it is important to note that the proposed model of $\textbf{G}$ has a value that it can be rigorously analyzed theoretically as all its components are well studied in graph theory literature and we take a union of independent graphs---see Subsection~\ref{sec:theory_motivation} for motivation for theoretical results.

\subsection{Distribution of Weights}\label{sec:splitting}

Parameter $\xi \in [0,1]$ controls the fraction of edges that are between communities; that is, it reflects the amount of noise in the network. Its role is similar to the role of the mixing parameter $\mu$ in the original \textbf{LFR} model.
We split weights $\textbf{w}$ into $\textbf{y}$ and $\textbf{z}$ as follows, keeping the same value of $\xi$ for each vertex (recall that $\textbf{y}$ will be associated with clusters and $\textbf{z}$ will be associated with the noise):
\begin{eqnarray*}
\textbf{y} = (y_1, \ldots, y_n) &=& (1-\xi) \textbf{w} = ((1-\xi) \cdot w_1, \ldots, (1-\xi) \cdot w_n), \\
\textbf{z} = (z_1, \ldots, z_n) &=& \xi \textbf{w} = (\xi \cdot w_1, \ldots, \xi \cdot w_n).
\end{eqnarray*}
However, in order to add more flexibility, one may allow different coefficients $\xi$ for each cluster. Let $(\xi_1, \ldots, \xi_k) \in [0,1]^k$. In the next subsection, vertices will be assigned into clusters: vertex $i\in[n]$ will be assigned to cluster $f(\sigma(i)) \in [k]$. Then,
\begin{eqnarray*}
\textbf{y} = (y_1, \ldots, y_n) &=&  ((1-\xi_{f(\sigma(1))}) \cdot w_1, \ldots, (1-\xi_{f(\sigma(n))}) \cdot w_n), \\
\textbf{z} = (z_1, \ldots, z_n) &=& (\xi_{f(\sigma(1))} \cdot w_1, \ldots, \xi_{f(\sigma(n))} \cdot w_n).
\end{eqnarray*}
This variant is important if one wants to mimic the original \textbf{LFR} model as closely as possible, that is, to try to keep the fraction of internal edges for each cluster equal; otherwise, using the same $\xi$ for all vertices suffice---see Subsection~\ref{sec:global_vs_local} for more discussion.

\subsection{Assigning Vertices into Clusters}\label{sec:assignment}

Our task now is to assign vertices into clusters, that is, to define the mapping $f : [n] \to [k]$ from vertices to clusters. Our goal is to design a fast algorithm that produces an assignment selected uniformly at random from some natural class of admissible assignments (formal definition is provided below).

The main problem is that vertices of large degree cannot be assigned to small clusters, as we aim to generate simple graphs for some applications of the proposed model. Recall that the weight vector $\textbf{w}$ will be split into $\textbf{y}$ and $\textbf{z}$ that will guide the process of generating cluster graphs and, respectively, the background graph. All edges within cluster graphs will end up between vertices of the same community. On the other hand, only some fraction of the background edges will be also present there as an effect of the random sampling. Unfortunately, the number of such edges depends on the mapping $f$ we are about to create, and so it is not known at this point. So how can we decide which vertex can be assigned to a given cluster leaving enough room for not only edges from the cluster graph but also for additional edges coming from the background graph? Fortunately, this ``chicken and egg'' problem can be solved as there exists a universal upper bound $x_i$ for $y_i$ that leaves enough room for the edges coming from the background graphs that works for all $i \in [n]$, namely,
\begin{equation}\label{eq:phi}
x_i := \Big\lceil (1 - \xi \phi ) w_i \Big\rceil,
\end{equation}
where $\phi := 1 - \sum_{\ell \in [k]} (s_{\ell}/n)^2$.

The reason for this choice of $x_i$ is as follows. In Subsection~\ref{sec:matching_global_mu}, we will show that the expected number of edges between communities is equal to $\xi \mu_0$, where $\mu_0=1-\sum_{\ell\in [k]} (W_{\ell}/W)^2$ ($W_{\ell}$ is the volume of cluster $\ell$, and $W$ is the volume of the whole graph). If vertices are assigned to clusters randomly, then the expected value of $W_{\ell}$ is $s_{\ell} W$. It follows that $\phi$ is a good approximation of $\mu_0$ that is not known at this point. In any case, since $\phi<1$, we observe that $x_i \ge (1-\xi \phi)w_i \ge (1-\xi)w_i = y_i$ and so there is definitely room for edges of the cluster graphs.

\medskip

Let us call an assignment of vertices into clusters \textbf{admissible} if each vertex $i \in [n]$ is assigned to cluster $j \in [k]$ with $x_i \le s_j-1$. Recall that our goal is to select one admissible assignment uniformly at random. This condition is a necessary condition for the existence of a simple graph that this cluster induces. Note that it is only a necessary condition; in fact, the corresponding degree sequence has to be graphic. (A \textbf{graphic} sequence is a sequence of numbers which can be the degree sequence of some graph; see, for example~\cite{West} or any other textbook on graph theory for more.) We use this slightly weaker condition because it is much simpler and more convenient to use which gives us an easier framework to work with. Finally, let us stress that for the final graph $\textbf{G}$ to be able to be simple, we get some additional constraints on admissible assignments of vertices into clusters. Not only $\textbf{z}$ and all $\textbf{y}_i$'s must be graphic but the union of all graphs needs to be simple as well. However, in practice, this causes no issue as we usually deal with sparse graphs that leave a lot of room for graphs to be fit.

Indeed, in practice the probability that a non-graphic degree sequence is obtained for some cluster graph is extremely low. However, in order to deal with such potential problematic situations the algorithm tries to assign as many edges as possible to stay within the cluster graph and move the remaining ones to the background graphs. See the end of Subsection~\ref{sec:two_models} for more details.

\medskip

A formal definition is slightly technical. Suppose that vertices are sorted according to their bounds on the expected/exact internal degree, that is, $x_1 \ge x_2 \ge \ldots \ge x_n$. Similarly, suppose that cluster sizes are sorted, that is, $s_1 \ge s_2 \ge \ldots \ge s_k$. In order to define the assignment of vertices into clusters, we need the following auxiliary sequence $s_{\le \ell}$. For each $\ell \in [k] \cup \{0\}$, let
$$
s_{\le \ell} := \sum_{i=1}^{\ell} s_i.
$$
In particular, $s_{\le 0}=0$ and $s_{\le k} = n$. Function $f:[n] \to [k]$, that we informally introduced earlier, is defined as follows. For each $i\in[n]$ and $j \in [k]$, we fix
$$
f(i) = j \qquad \text{if and only if} \qquad s_{\le j-1} < i \le s_{\le j}.
$$
The \textbf{assignment} now can be viewed as a permutation $\sigma:[n] \to [n]$---vertex $i \in [n]$ is assigned to cluster $f(\sigma(i)) \in [k]$. Such assignments guarantee that the right number of vertices is assigned to each cluster but vertices of large degree could be assigned to small clusters. Let $\mathcal{A}$ be the set of \textbf{admissible assignments} defined as follows:
$$
\mathcal{A} := \Big\{ \sigma:[n] \to [n] : x_i \le s_{f(\sigma(i))} - 1 \textrm{ for all } i \in [n] \Big\}.
$$
In other words, no vertex in admissible assignment gets assigned to a cluster of size smaller than or equal to its expected/exact degree. Our goal is to select one member of the family $\mathcal{A}$ uniformly at random.

Sampling with uniform distribution is often a difficult task. Of course, generating one permutation with uniform distribution on the set of \emph{all} permutations is easy and can be done in many different ways. If such permutation falls into $\mathcal{A}$, then we could accept it; otherwise, we repeat the process until we get one that does it. Unfortunately, the size of $\mathcal{A}$ comparing to $n!$, the number of all possible permutations, can be very small so this rejection sampling process is not feasible from practical point of view. However, this point of view does have theoretical implications and might be useful in the future for analyzing the model.

Fortunately, sampling uniformly from $\mathcal{A}$ turns out to be relatively easy. To that end, we will use the following natural algorithm. (See also a pseudo-code in the Appendix.) Recall that vertices are sorted according to their bounds on internal degrees, that is, sequence $\textbf{x}=(x_1, \ldots, x_n)$ is non-increasing. Consider vertices, one by one, starting with vertices that are associated with large values of $x_i$, and assign them randomly to a cluster that has size larger than the corresponding bound and still has some ``free spots''; that is, a cluster of size $s_j$ is considered for a vertex of degree $x_i$ if $x_i \le s_j-1$ and the number of vertices already assigned to it is less than $s_j$. The probability that a given vertex is assigned to a given cluster is proportional to the number of ``free spots'' that remain in that cluster.

The reason why this algorithm produces an admissible assignment uniformly at random comes from the fact that clusters that are assigned to earlier vertices could also be assigned to vertices considered later. In other words, it is \emph{not} the case that vertices considered earlier could make decisions that create more (or less) choices for vertices considered later. They need to be assigned somewhere and, regardless of where they get assigned, the number of choices left for future vertices is not affected. In particular, the algorithm always terminates, \emph{unless} $\mathcal{A} = \emptyset$.

To see a formal argument, let
$$
t_i := \max \{ s_{\le \ell} : x_i \le s_{\ell}-1 \}.
$$
It is straightforward to see that $\sigma \in \mathcal{A}$ \emph{if and only if} $\sigma(i) \in [t_i]$ for all $i \in [n]$. Note that, for any given admissible permutation $\sigma \in \mathcal{A}$, our algorithm produces it with probability $p$ that is only a function of $t_i$ but does not depend on $\sigma$. Indeed, it is easy to see that
$$
p = \prod_{i=1}^n \frac {1}{t_i-i+1},
$$
as there are $t_i-(i-1)$ available ``free spots'' for a vertex $i\in[n]$. Clearly, the algorithm does not produce any permutation that is not admissible. Hence, indeed, the algorithm generates a permutation from $\mathcal{A}$ uniformly at random. As mentioned above, the algorithm fails only if $\mathcal{A} = \emptyset$.

\subsection{Exact vs.\ Expected Degree Distribution --- Two Variants of the Model}\label{sec:two_models}

We will consider two variants of the model: the first one generates graphs with the expected degree distribution $\textbf{w}$ (related to the well-known Chung-Lu model), and the second one with the exact degree distribution $\textbf{w}$ (related to another well-known model, the configuration model). We will start with the description of the first variant, as it is slightly easier. However, it is presumably the case that the practitioners prefer the second variant. (In particular, a potential appearance of isolated vertices in sparse Chung-Lu models might not be desirable for practical purposes.)

\medskip

Recall that at this point we have vertices assigned to clusters: vertex $i \in [n]$ belongs to cluster $f(\sigma(i))$. Moreover, the weight vector $\textbf{w}$ is split into two vectors $\textbf{y}$ and $\textbf{z}$ that will guide the creation of cluster graphs $G_i$ ($i \in [k]$) and, respectively, the background graph $G_0$. We need to specify how we actually do it and how we deal with potential problems after taking the union of these graphs.

\subsubsection{The Expected Degree Distribution}

In this variant of the model, we use the Chung-Lu model that produces a random graph with expected degree sequence following a given sequence.

\subsubsection*{Chung-Lu Model}

Let $\textbf{w}=(w_1, \ldots, w_n)$ be any vector of $n$ real numbers, and let $W=\sum_{i=1}^n w_i$. We define $\mathcal{C}(\mathbf{w})=([n],E)$ to be the probability distribution of graphs on the vertex set $[n]$ following the well-known Chung-Lu model~\cite{CL2006, Seshadhri2012, Kolda2014, Winlaw2015}. In this model, each set $e=\{i,j\}$, $i,j \in [n]$, is independently sampled as an edge with  probability given by:
\[
\Pro(i,j) =
\begin{cases}
\frac{w_i w_j}{W}, & i \ne j \\
\frac{(w_i)^2}{2W}, & i = j.
\end{cases}
\]
(Let us mention about one technical assumption. Note that it might happen that $\Pro(i,j)$ is greater than one and so it should really be regarded as the expected number of edges between $i$ and $j$; for example, as suggested in~\cite{Newman_book}, one can introduce a Poisson-distributed number of edges with mean $\Pro(i,j)$ between each pair of vertices $i$, $j$. However, since typically the maximum degree $\Delta$ satisfies $\Delta^2 \le 2 |E|$ it rarely creates a problem and so we may assume that $\Pro(i,j) \le 1$ for all pairs.)

One desired property of this random model is that it yields a distribution that preserves the expected degree for each vertex, namely: for any $i \in [n]$,
$$
\E[\deg(i)] = \sum_{j \in [n] \setminus \{i\}} \frac{w_i w_j}{W} + 2 \cdot \frac{(w_i)^2}{2W} = \frac{w_i}{W} \sum_{j \in [n]} w_j = w_i.
$$

\subsubsection*{Theoretical Approach}

The original Chung-Lu model is a multi-graph so it is natural and convenient to stay with multi-graphs in our model too. We simply take $G_i = \mathbf{G}(\textbf{y}_i) = \mathcal{C}(\textbf{y}_i)$ for each $i \in [k]$, and $G_0 = \mathbf{G}(\textbf{z}) = \mathcal{C}(\textbf{z})$.

\subsubsection*{Practical Approach --- Insisting on Simple Graphs}

From practical point of view, it is desired to generate a simple graph and use a fast algorithm that does it. In order to achieve both things, we use a version of the (fast) Chung-Lu model that produces the graph with a given number of edges. As a result, we need to round some numbers to integers. We use the following randomized way that is also used in the original \textbf{LFR} model. For a given integer $k \in \Z$ and real number $\ell \in [0,1)$, let
$$
\Big\lfloor k+\ell \Big\rceil =
\begin{cases}
k & \textrm{ with probability } 1-\ell \\
k+1 & \textrm{ with probability } \ell.
\end{cases}
$$
(Note that the expected value of random variable $\lfloor k+\ell \rceil$ is equal to $k+\ell$.)

We independently generate cluster graphs $G_i$ ($i \in [k]$) as follows. Note that $\sum_{v \in V_i} y_v/2$ is the expected number of edges in $\mathcal{C}(\textbf{y}_i)$. We fix
$$
e_i := \left\lfloor \frac 12 \sum_{v \in V_i} y_v \right\rceil \in \N \cup \{0\},
$$
and then we generate the Chung-Lu graph $\mathcal{C}(\textbf{y}_i)$ conditioning on not having parallel edges or loops, and having exactly $e_i$ edges. This can be done in a fast way. We independently sample two vertices $i$ and $j$ with probabilities proportional to their weights. If $i \neq j$ and adding an edge $\{i,j\}$ does not create a parallel edge, then we accept it. We continue this process until $e_i$ edges are created.

Once all cluster graphs are created, we move to the background graph. In order to keep the total number of edges as desired, we fix
$$
e := \frac 12 \sum_{v \in V} w_v - \sum_{i=1}^k e_i.
$$
Note that $\sum_{v \in V} w_v$ is usually an even integer (since vector $\textbf{w}$ corresponds to the degree sequence) so $e \in \N \cup \{0\}$. (If not, we may replace $\sum_{v \in V} w_v/2$ with $\lfloor \sum_{v \in V} w_v/2 \rceil$.) Note also that the expected value of $e$ is equal to $\sum_{v \in V} z_v/2$, the expected number of edges in $\mathcal{C}(\textbf{z})$. We generate the Chung-Lu graph $\mathcal{C}(\textbf{z})$ conditioning on not having loops, not creating parallel edges (in the union of all cluster graphs and the background edges created so far!), and having exactly $e$ edges. To that end, we use the same fast algorithm as before. Note that, as long as the whole graph is sparse (which is typically the case), the second step is fast since not too many collisions occur, even if some of the cluster graphs $G_i$ ($i \in [k]$) are dense.

\subsubsection{The Exact Degree Distribution}

This variant of the model uses the configuration model (instead of the Chung-Lu model) that produces a random graph with a given degree sequence. However, this change brings a few small issues that need to be dealt with.

\subsubsection*{Configuration Model}

Let $\textbf{w}=(w_1, \ldots, w_n)$ be any vector of $n$ non-negative integers such that $W:=\sum_{i=1}^n w_i$ is even. We define a random miuti-graph $\mathcal{M}(\textbf{w})$ with a given degree sequence known as the \textbf{configuration model} (sometimes called the \textbf{pairing model}), which was first introduced by Bollob\'{a}s~\cite{bollobas2}. (See~\cite{Bender_Canfield, Wormald} for related models and results.)

Let us consider $W$ \textbf{configuration points} partitioned into $n$ labelled buckets $v_1,\ldots,v_n$; bucket $v_i$ consists of $w_i$ points. A \textbf{pairing} of these points is a perfect matching into $W/2$ pairs. (There are $W! / ((W/2)! 2^W)$ such pairings.) Given a pairing $P$, we may construct a multi-graph $G(P)$, with loops and parallel edges allowed, as follows: the vertices are the buckets $v_1,\ldots, v_n$, and a pair $\{x,y\}$ in $P$ corresponds to an edge $\{v_i,v_j\}$ in $G(P)$ if $x$ and $y$ are contained in the buckets $v_i$ and $v_j$, respectively. We take a pairing $P$ uniformly at random from the family of all pairings of $W$ points and set $\mathcal{M}(\textbf{w}) = G(P)$.

It is an easy but a fundamental fact that the probability of a random pairing corresponding to a given simple graph $G$ is independent of the graph. Indeed, an easy calculation shows that every simple graph corresponds to exactly $\prod_{i=1}^n w_i!$ pairings. Hence, the restriction of the probability space of random pairings to simple graphs is precisely $\mathcal{S}(\textbf{w})$, the uniform probability space of all \emph{simple} graphs with a given degree sequence. Moreover, it is well known that if
$$
\sum_{i=1}^n w_i = \Theta(n) \qquad \textrm{ and } \qquad \sum_{i=1}^n w_i^2 = O(n),
$$
then the expected number of loops and multiple edges that are present in $\mathcal{M}(\textbf{w})$ is $O(1)$ and so the probability that $\mathcal{M}(\textbf{w})$ is simple tends to $\delta=\delta(\textbf{w}) > 0$ which depends on $\textbf{w}$ but is always separated from zero. As a result, event holding a.a.s.\ (that is, with probability tending to 1 as $n \to \infty$) over the probability space $\mathcal{M}(\textbf{w})$ also holds a.a.s.\ over the corresponding space $\mathcal{S}(\textbf{w})$. For this reason, asymptotic results over random pairings immediately transfer to $\mathcal{S}(\textbf{w})$. One of the advantages of using this model is that the pairs may be chosen sequentially so that the next pair is chosen uniformly at random over the remaining (unchosen) points.

\subsubsection*{Distribution of Weights}

We assume that integer-valued vector $\textbf{w}$ is such that $\sum_i w_i$ is even so that a given degree sequence is feasible. (As mentioned earlier, it is only a trivial, necessary condition---in fact, $\textbf{w}$ should be a graphic sequence.) Recall that the weights, vector $\textbf{w}$, is split into real-valued vectors $\textbf{y}$ and $\textbf{z}$. However, since we deal with exact degree sequences not expected ones, this time we have two additional constraints that we need to satisfy, namely, that
a) all involved weights are integers,
and
b) for each of the $k$ clusters, the corresponding sum of weights is even.
Note that once these conditions are satisfied for all cluster graphs, the background graph immediately has them too---all degrees are integers and the sum of weights is even.

We split $\textbf{w}$ into integer-valued vectors $\hat{\textbf{y}}=(\hat{y}_1, \ldots, \hat{y}_n)$ and $\hat{\textbf{z}}=(\hat{z}_1, \ldots, \hat{z}_n)$ as follows. For each cluster $i \in [k]$, we identify the \textbf{leader}, vertex of the largest weight in cluster $i$. (If more than one vertex has the largest weight, we select one of them to be the leader, arbitrarily.) In order to deal with non-integer values, for all vertices $i \in [n]$ that are \emph{not} leaders, we set $\hat{y}_i = \lfloor y_i \rceil$. For the remaining $k$ vertices, the leaders, we round $y_i$ up or down so that the sum of weights in each cluster is even. (If some leader has the weight $y_i \in \N$  and the sum of weights in its cluster is odd, then we randomly make a decision whether subtract or add 1 to make the sum to be even.)

\subsubsection*{Theoretical Approach}

We take $G_i = \mathbf{G}(\hat{\textbf{y}}_i) = \mathcal{M}(\hat{\textbf{y}}_i)$ for each $i \in [k]$, and $G_0 = \mathbf{G}(\hat{\textbf{z}}) = \mathcal{M}(\hat{\textbf{z}})$. Some of the involved graphs might not be simple but the expected number of loops and parallel edges is small, especially for sparse graphs. We have a few options how to deal with them.
The first option is the easiest: we could do nothing and work with multi-graphs.
Alternatively, we could condition on all $G_i$ ($i\in [k] \cup \{0\}$) to be simple. From theoretical point of view, this model is equally easy to analyze, provided that for each $G_i$, the probability of getting a simple graph tends to a constant as $n\to \infty$ (does not matter how small it is and could be different for each $i \in [k] \cup \{0\}$). It is known that under some mild assumptions this is the case (in particular, the order of each cluster graph should tend to infinity with $n$, etc.)---see above for the discussion around the configuration model. Let us remark that even though all $G_i$'s are simple, it is not guaranteed that the final graph, $\textbf{G}$, is simple as edges from $G_0$ can overlap with edges of $G_i$ for some $i \in [k]$. Hence, we could condition on $\textbf{G}$ to be simple. Unfortunately, this model might be more challenging to analyze (as it introduces some dependencies between the background graph and the cluster graphs) but this is certainly worth investigating in the future work.

\subsubsection*{Practical Approach --- Insisting on Simple Graphs}

Before we discuss how we apply these observations to our problem, let us discuss a general approach and some theoretical, asymptotic results. Let us generate a random graph with a given degree sequence using the configuration model. If it happens that it is a simple graph, it is a uniformly distributed random graph from the family of simple graphs with this degree sequence. Suppose then that it is not simple. It is known that after performing some kind of ``switching'' we get a random graph that is very close to the uniform distribution and we should solve all problems in $O(1)$ time. Indeed, in~\cite{Janson}, it is proved that, assuming essentially a bounded second moment of the degree distribution, the configuration model with the simplest types of switchings yields a simple random graph with an almost uniform distribution,  in  the  sense  that  the  total  variation distance is $o(1)$. For each parallel edge $uv$, one needs to choose a random edge $xy$, remove $uv$, $xy$, and with probability $1/2$ add $ux$, $vy$; otherwise, add $uy$, $vx$.

Let us now explain how we actually apply switchings to our problem. We start with the configuration model to generate cluster multi-graphs $G_i$ ($i \in [k]$). We then apply switchings to get a family of simple graphs.

After that, we use the configuration model to generate the background graph $G_0$ and use switchings to remove loops and parallel edges.

After taking the union, more parallel edges could be created. As usual, we use switchings to remove them. However, this time we restrict ourselves to edges in the background graph and switch only those. This can be done since all graphs $G_i$ are simple at this point and so collisions must involve at least one edge from the background graph. During switching more collisions can be created but each collision again involves at least one edge from the background graph. We do this to preserve the number of internal edges within cluster; the cluster graphs are not affected by this final round of switchings.

In order for our algorithm to be fast in all potential situations, we have implemented a procedure that controls the process of fixing multiple edges and self loops so that it is not extremely slow (and, in particular, to be robust against a mentioned earlier rare possibility of obtaining a non-graphic degree sequence). If some cluster graph $G_i$ is extremely dense, it might be computationally expensive (or simply impossible for non-graphic degree sequence) to sample a correct replacement that maintains all the desired constraints. This situation is extremely rare but if it happens, then we retry it only for a limited number of times. This creates a small bias for the number of edges captured within that community, but we have empirically found that it happens less than 1 per 1,000,000 edges for typical tight configurations of the model so the bias should not be noticeable in practice. It is possible to resolve all conflicts exactly (unless, of course, a non-graphic degree sequence is obtained, which is rare) so this is simply a trade-of between the speed and the quality of the implementation.

In summary, the conflict resolution algorithm we use for each cluster graph $G_i$ works as follows:
\begin{enumerate}
\item perform a standard configuration model on $G_i$ but put all self loops and multiple edges in a \emph{recycle} list assigned to this graph;
\item iteratively, remove one edge from the \emph{recycle} list and try to rewire it with randomly selected edge from $G_i$ including those from the \emph{recycle} list; this process is tried as many times as the target number of edges in $G_i$ (so, in expectation, each edge is tried for rewiring once); if we successfully do the switching, then we move forward; otherwise, we return the chosen edge back to the \emph{recycle} list;
\item the whole process is repeated as long as we are able to find a good rewiring for an edge in \emph{recycle} until \emph{recycle} becomes empty or the number of times we were unable to reduce the \emph{recycle} list size is equal to the size of \emph{recycle}, that is, we unsuccessfully tried to recycle all edges in \emph{recycle}; in such a case, we give up and move the remaining degrees of the vertices forming those unmatched edges from $G_i$ to the global graph so that the final degree of all vertices in the union graph follows $\mathbf{w}$---as noted above, this action is extremely rare---approximately less frequent than once per 1,000,000 edges.
\end{enumerate}
For the background graph we follow the same procedure. However, we do not ``give up'' recycling and follow the process until all required edges are created. As the background graph is sparse, this process is very fast in practice.

\section{Comparing \textbf{ABCD} and \text{LFR}}\label{sec:comparison}

The role of parameter $\xi$ in \textbf{ABCD} is similar to the one of parameter $\mu$ in \textbf{LFR}; however, they are not the same! If $\xi=\mu=0$, then in both models all edges are within communities, but if $\xi=\mu=1$, then \textbf{ABCD} is a random graph and so is substantially different than \textbf{LFR} which produces ``anti-communities''. As a result, in order to compare the two models one needs to tune the parameters such that the corresponding densities of communities are comparable.

The are two natural ways of distributing the weight $\textbf{w}$---the first one preserves the densities globally whereas the second one preserves it locally for each vertex in the graph. We will independently consider both approaches. After that we will discuss the difference between the two and their implications. However, before that let us recall one subtle caveat we already discussed in Section~\ref{sec:assignment} when we assigned vertices into clusters.

If one creates a pure \textbf{ABCD} graph, then $\xi$ is known upfront and there is no issue. Now, we try to find $\xi$ for \textbf{ABCD} that matches given $\mu$ for \textbf{LFR}. The problem is that we cannot compute $\xi$ before vertices are assigned to clusters. On the other hand, to do the assignment we need to bound the number of neighbours of each vertex that belong to its own cluster graph that is a function of $\xi$---recall equation~(\ref{eq:phi}). To overcome this ``chicken and egg'' problem, we apply some universal upper bound $x_i$ for $y_i$, namely $x_i:=\lceil (1-\mu)w_i\rceil$ to do the assignment, and then compute $\xi$. Hence, in what follows we may assume that the assignment is given to us and we simply tune $\xi$ to match given $\mu$.

\subsection{Recovering the Mixing Parameter (Globally)}
\label{sec:matching_global_mu}

In this scenario, we start with a fixed $\xi$ that will be applied for all vertices regardless to which cluster they belong to. Recall that $V_\ell = \{t \in V : f(\sigma(t)) = \ell\}$ ($\ell \in [k]$) is the set of vertices assigned to cluster $\ell$.  Let $W=\sum_{t \in V} w_t$ be the volume of $G$, and let $W_{\ell} = \sum_{t \in V_\ell} w_t$ be the expected/exact volume of vertices of cluster $\ell$. Clearly, $W = \sum_{\ell \in [k]} W_{\ell}$.

There are two models (namely, Chung-Lu and Configuration Model) used to generate multi-graphs $G_i$ ($i \in [k] \cup \{0\}$) but both of them have the property that edges occur with probability proportional to the product of the weights of the two endpoints. Consider two vertices $i,j$ with weights $w_i$ and, respectively, $w_j$. If they are in different clusters ($f(\sigma(i)) \neq f(\sigma(j))$), then the probability that they are adjacent is equal to
$$
\frac {z_i z_j}{\sum_{t\in V} z_t} = \frac {\xi w_i \cdot \xi w_j}{\sum_{t \in V} \xi w_t} = \xi \frac {w_i w_j}{\sum_{t \in V} w_t} = \xi \frac {w_i w_j}{W}.
$$
(In fact, for multi-graphs it is the expected number of edges as the value above could potentially exceed 1. However, it is a rare situation in practice.)
It follows that the fraction of edges that are between communities is equal to
\begin{eqnarray*}
\frac {1}{W} \sum_{i \in V} \sum_{j \in V \setminus V_{f(\sigma(i))}} \xi \frac {w_iw_j}{W} &=& \frac {\xi}{W^2} \sum_{i \in V} w_i \sum_{j \in V \setminus V_{f(\sigma(i))}} w_j = \frac {\xi}{W^2} \sum_{i \in V} w_i (W - W_{f(\sigma(i))}) \\
&=& \frac {\xi}{W^2} \sum_{\ell \in [k]} W_{\ell} (W - W_{\ell}) = \frac {\xi}{W^2} \left( W^2 - \sum_{\ell \in [k]} W_{\ell}^2 \right) \\
&=& \xi \left( 1 - \sum_{\ell \in [k]} (W_{\ell}/W)^2 \right) = \xi \ \mu_0,
\end{eqnarray*}
where $\mu_0 := 1 - \sum_{\ell \in [k]} (W_{\ell}/W)^2$.
Hence, in order to mimic the structure of the \textbf{LFR} graph, one should consider

\begin{equation}
\xi = \frac {\mu}{\mu_0} = \mu \left( 1 - \sum_{\ell \in [k]} (W_{\ell}/W)^2 \right)^{-1}.
\label{eq:xi}
\end{equation}

On the other hand, if vertices $i$ and $j$ are in the same cluster $\ell$, the probability is equal to
\begin{eqnarray*}
\frac {z_i z_j}{\sum_{t \in V} z_t} + \frac {y_i y_j}{\sum_{t \in V_\ell} y_t} &=& \xi \frac {w_i w_j}{\sum_{t \in V} w_t} + (1-\xi) \frac {w_i w_j}{\sum_{t \in V_\ell}w_t} \\
&=& \frac {\xi w_i w_j}{W} + \frac{(1-\xi) w_i w_j}{W_l} = \frac {w_i w_j}{W} + (1-\xi) w_i w_j \left(\frac{W-W_\ell}{W \cdot W_l}\right).
\end{eqnarray*}
The expected number of neighbours of $i$ that are in cluster $\ell$ then equal to
\begin{eqnarray}
\sum_{j \in V_{\ell}} \left( \frac {\xi w_i w_j}{W} + \frac{(1-\xi) w_i w_j}{W_l} \right) &=& w_i \left( \xi \frac {W_\ell}{W} + (1-\xi) \right) = w_i \left( \frac {W_\ell}{W} + (1-\xi) \frac{W-W_{\ell}}{W} \right).
\label{eq:expected}
\end{eqnarray}

\medskip

Let us make one remark. Note that if $\mu > \mu_0$, then the corresponding value of $\xi$ is grater than 1. As a result, we cannot generate our random graph. One can see it as a potential problem but, in fact, it is the opposite. Such values of $\mu$ correspond to models in which the density between clusters is larger than the internal density. As discussed in Subsection~\ref{sec:ill-defined}, we should not be ever concerned with such networks with ``anti-communities''.

\subsection{Recovering the Mixing Parameter for Each Vertex (Locally)}

In this scenario, we consider a sequence of parameters $\xi_i$ ($i \in [k]$), one per each cluster. In the original LFR model, once the degree sequence $\textbf{w}=(w_1,\dots,w_n)$ is fixed, the algorithm tries to re-wire the edges such that for each vertex $i$, the internal degree is close to $(1-\mu) w_i$. There is some variability in the final ``local'' mixing parameters, but mainly due to the presence of low degree vertices which clearly must deviate from the desired ratio.

It is not clear if matching local parameters is what we want (see the discussion in the introduction and in Subsection~\ref{sec:global_vs_local}) but here is a possible way to modify the approach presented above in order to have local mixing parameters close to $\mu$. Instead of using the same ratio $\xi$ for splitting weights into background and cluster portions, one can carefully tune it and use different values of $\xi$ for different clusters.
Consider vertex $i$ with degree $w_i$ that belongs to a cluster with the total weight equal to $W_{f(\sigma(i))}$.
For the background graph $G_0=\textbf{G}({\bf z})$, let $z_i = \xi_{f(\sigma(i))} \cdot w_i$ be such that
$$
z_i \left(\frac{W-W_{f(\sigma(i))}}{W}\right) = w_i \cdot \mu. $$
Indeed, this is desired as only the $(W-W_{f(\sigma(i))})/W$ fraction of the background edges are expected to be present between the communities. It follows that the ratio for cluster $\ell \in [k]$ should be defined as follows:

\begin{equation}
\xi_{\ell} = \mu \left(\frac{W}{W-W_{\ell}}\right).
\label{eq:xi_i}
\end{equation}
As a result, for the cluster graph $G_{f(\sigma(i))} = \textbf{G}({\bf y}_{f(\sigma(i))})$ corresponding to the cluster of vertex $i$, we let $y_i = w_i - z_i = (1-\xi_{f(\sigma(i))}) \cdot w_i.$

\medskip

As before, there exists a threshold $\mu_1$ such that if $\mu > \mu_1$, then some value of $\xi_{\ell}$ is grater than 1 and so the model cannot be applied. This time
$$
\mu_1 = \min_{\ell \in [k]} \frac {W-W_{\ell}}{W} = 1 - \frac {\max_{\ell \in [k]} W_{\ell}}{W}.
$$

\subsection{The Comparison Between the Two Variants (Global vs.\ Local)}\label{sec:global_vs_local}

Let us summarize the difference between the two approaches discussed above. Both of them preserve the same number of edges between clusters: $\mu$ fraction of all edges are of this type (global property). The difference is how we split the degree of each vertex into internal degree and external one (local property). The original \textbf{LFR} model insists on each vertex keeping the same fraction of internal neighbours and the local version of our model (with $k$ parameters $\xi_i$) does this too. As a result, small clusters will be much denser than large clusters. Is it what we expect to happen in complex networks?

Suppose that two researchers have the same number of friends (say, 100) but belong to different communities. The first one, Bob, belongs to a small community (say, he is a mathematician doing some esoteric part of mathematics), the second one, Alice, is part of a large community (say, she is a data scientist). Suppose that 30\% of friends of Alice do data science. Should we expect 30\% of friends of Bob to be in his field? We believe the answer is no. It might be the case that there are less than 30 people around the world working on this subject! Coming back to the model, it seems that it makes more sense for the number of internal neighbours of a given vertex to be a function of the size of the cluster this vertex belongs to. As long as the probability that a given vertex is connected to another member of its cluster is larger than the probability of being adjacent to a random vertex in the whole graph, this vertex is a legit member of this cluster. This is what we propose in our first variant, the global version of our model (with only one parameter $\xi$).


\section{Experimental Results}\label{sec:experiments}

In this Section, we compare \textbf{ABCD} and \textbf{LFR} benchmarks with respect to their respective mixing parameters (Subsection~\ref{sec:exp_mixing_par}), the efficiency of the algorithms (Subsection~\ref{sec:speed}), and properties of the graphs they generate (Subsection~\ref{sec:exp_properties}). In order to perform fair comparisons, we fix the {\bf LFR} mixing parameter $\mu$ and then derive the corresponding parameters for {\bf ABCD}: $\xi$ via equation~(\ref{eq:xi}) for the global model, or $\xi_i$'s via equation~(\ref{eq:xi_i}) for the local model.

Instructions how to reproduce Figures~\ref{fig:reg} and~\ref{fig:properties} can be found on-line\footnote{\texttt{https://github.com/bkamins/ABCDGraphGenerator.jl/tree/master/instructions}}. We do not make the codes for producing the exact results given in Section~\ref{sec:speed} public, as they required some technical changes in comparison to publicly available implementations of the algorithms; in particular, we wanted to measure only the graph generation time without saving it to disk. Using the instructions presented on GitHub, that are based on end-user versions of codes, allow to reproduce the presented results with high accuracy while minimizing complexity of execution of the experiments. Moreover, for Figures~\ref{fig:reg} and~\ref{fig:properties} we performed a slightly more exact comparison than presented on GitHub; that is, the same vertex degrees and community sizes are provided to all algorithms rather than generating them independently each time to make sure that the corresponding graph generation processes are compared on exactly the same data. However, the results are very similar to what is obtained with the simplified approach available on-line.
The modified implementations that were used to generate figures in this paper can be made available upon request. 

\subsection{Global vs.\ Local Mixing Parameters}\label{sec:exp_mixing_par}

We showed above that in the {\bf ABCD} model (global variant), we expect a larger proportion of internal edges for larger communities. This implies a negative correlation between the community-wise mixing parameters $\mu_i$ (that is, the proportion of external edges for a given community) and the community sizes $s_i$. This is slightly different than in the {\bf LFR} model which tries to preserve the same community-wise $\mu_i$ for each community. We showed that the {\bf ABCD} model can be easily modified to mimic this property by defining community-wise parameters $\xi_i$, which we refer to as the local variant of the model.

\begin{figure}[ht]
\begin{center}
\includegraphics[width=5.1cm]{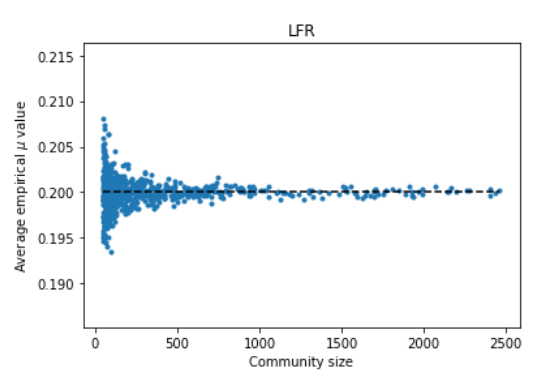}
\hspace{0.1cm}
\includegraphics[width=5.1cm]{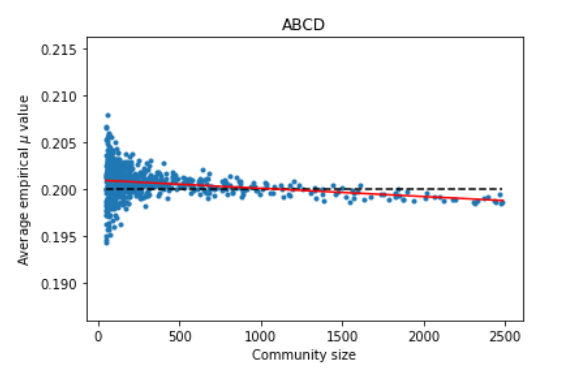}
\hspace{0.1cm}
\includegraphics[width=5.1cm]{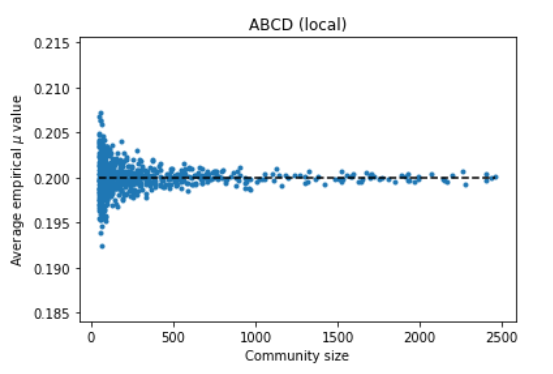}
\end{center}
\caption{Comparing the behaviour of graphs with $n=250,000$ vertices generated from 3 models: {\bf LFR}, {\bf ABCD} (with the configuration model), and its local variant. We used the same degree and community sizes distributions obtained with parameters: $\bar{w}=25$, $w_{max}=1500$ and $\gamma=2.5$ for the degrees, and $c_{min}=50$, $c_{max}=2500$ and $\beta=1.5$ for the community sizes.
We see that with {\bf LFR} and {\bf ABCD} (local variant), the expected community-wise mixing parameter $\mu_i$ is constant while for the {\bf ABCD} model, it decreasing as a function of the community size.
}
\label{fig:reg}
\end{figure}

In Figure~\ref{fig:reg}, we illustrate this behaviour for graphs with $n=250{,}000$ vertices and the same degree and community sizes distributions. We plot the mixing parameter $\mu_i$ for each community as a function of its size. For comparison purpose, the dashed black horizontal line corresponds to the constant value $\mu=0.2$.
For the global variant of the {\bf ABCD} model, we also display the regression line obtained by fitting the expected values for the $\mu_i$ using the formula~(\ref{eq:expected}).
In each case, due to rounding issue we see more variability for small communities, as expected. For {\bf LFR}, we see that the average value stays close to $\mu=0.2$ while with the global variant of the {\bf ABCD}, the value decreases with the community size matching the expected behaviour quite well. Using the local version of {\bf ABCD}, we see that we get similar behaviour to the {\bf LFR} model.

\subsection{Efficiency Comparison}\label{sec:speed}

\begin{figure}[ht]
\begin{center}
\includegraphics[width=5.1cm]{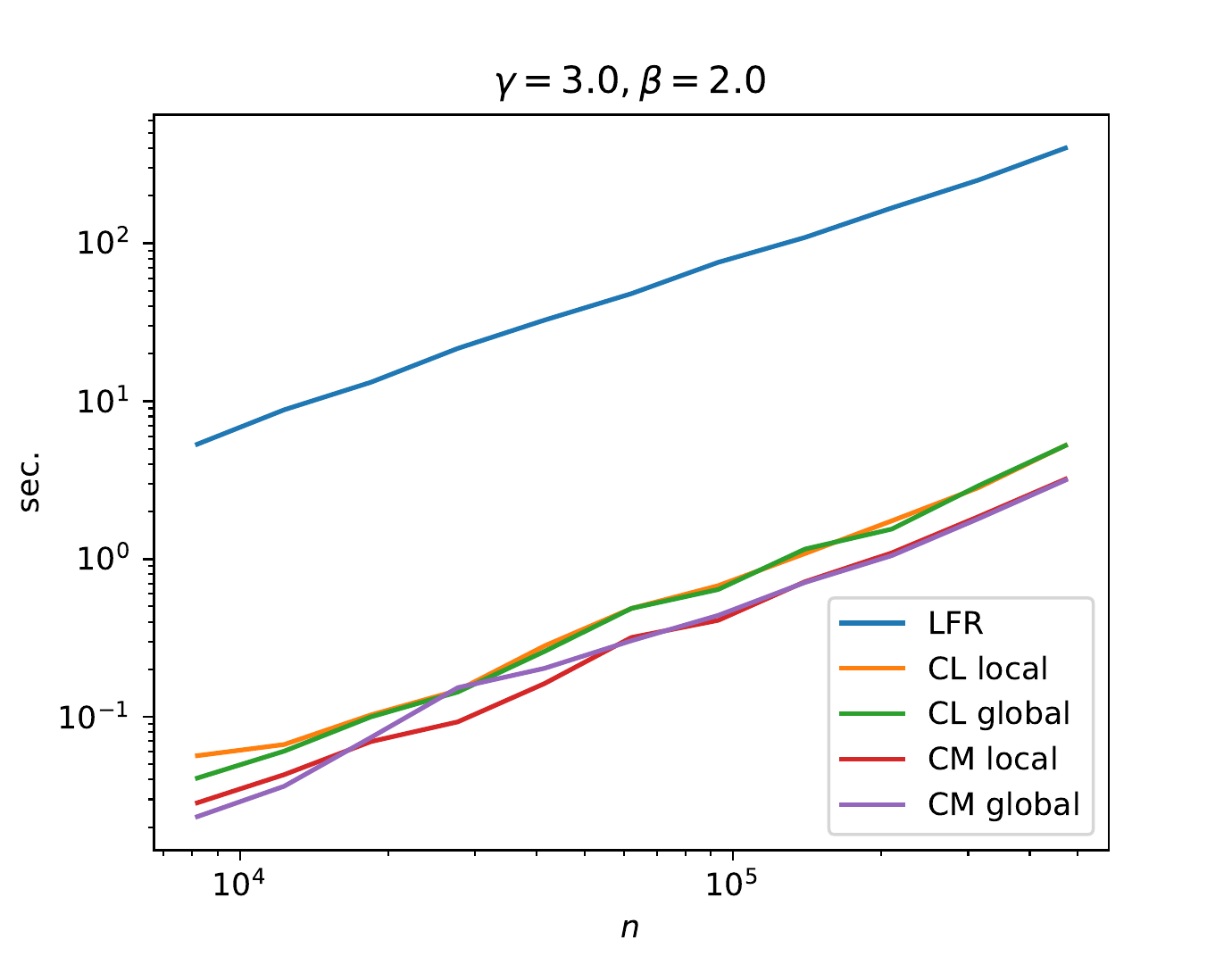}
\hspace{0.1cm}
\includegraphics[width=5.1cm]{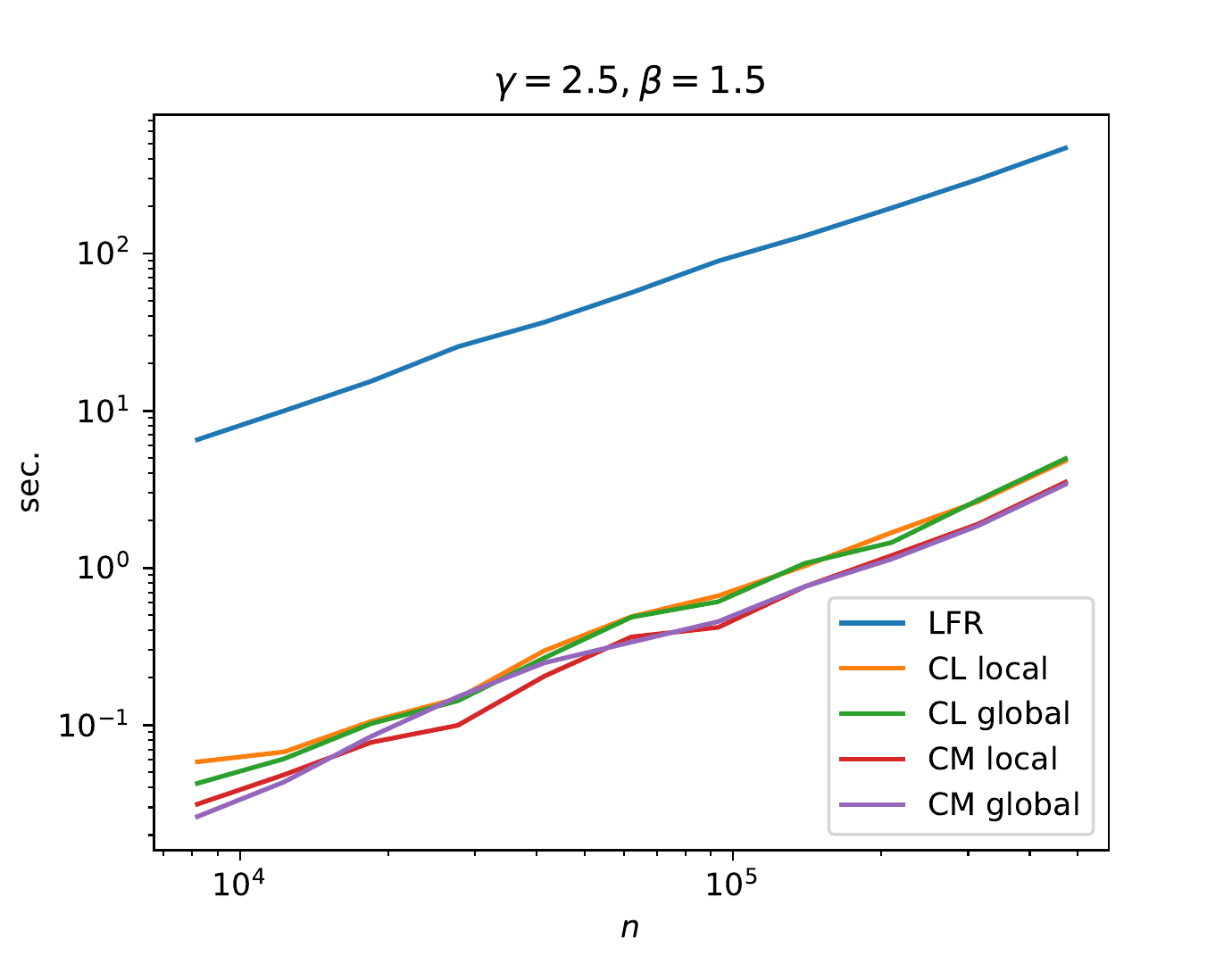}
\hspace{0.1cm}
\includegraphics[width=5.1cm]{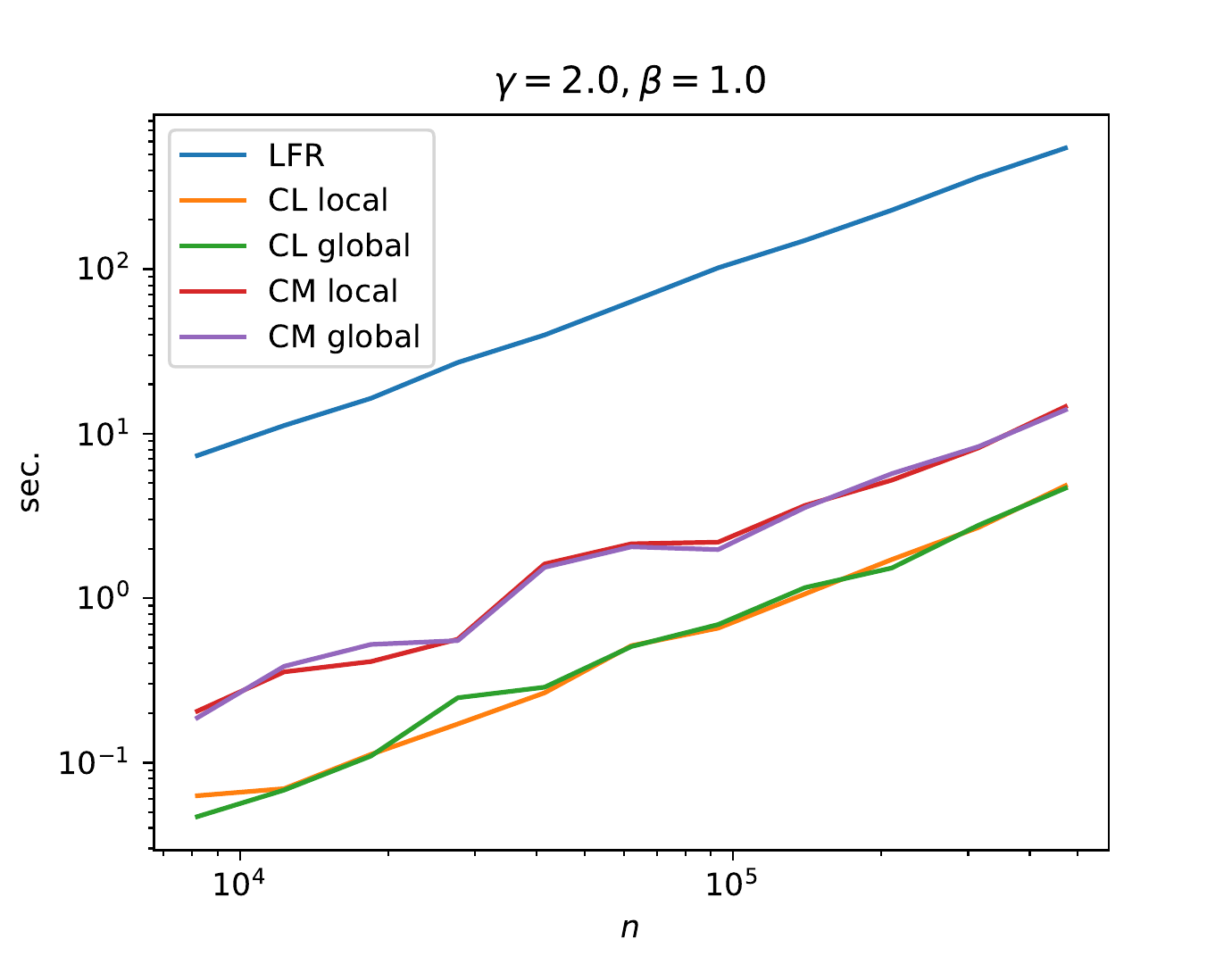}
\end{center}
\caption{Generation times in seconds of the {\bf LFR} and the {\bf ABCD} models; CL indicates the Chung-Lu model and CM indicates the configuration model.} 
\label{fig:timing}
\end{figure}

In this subsection, we compare efficiency of the generating algorithms. All the results were obtained on a single thread of Intel Core i7-8550U CPU @ 1.80GHz, run under Microsoft Windows 10 Pro, and performing all computations in RAM. The computations for {\bf LFR} were performed using the C++ language implementation\footnote{\texttt{https://github.com/eXascaleInfolab/LFR-Benchmark\_UndirWeightOvp/}}.
For {\bf ABCD}, the Julia 1.3 language implementation was used \cite{juliabezanson} in order to ensure high performance of graph generation, while keeping the size of the code base small. We tested all four combinations of the \textbf{ABCD} model (Chung-Lu vs.\ Configuration Model, and global vs.\ local variant).

In order for comparison to be fair, we first generated the degree distribution and the distribution of cluster sizes, and then used it for all five algorithms tested (\textbf{LFR} and 4 combinations of \textbf{ABCD}). Only the times to generate the corresponding graphs (single threaded, in memory, without storing the outcome on a hard drive) were measured, assuming the degree distribution and the distribution of cluster sizes are given---these steps are fast anyway.

\medskip

The models were generated for $\mu=0.2$ (and its counterparts for $\xi$'s for \textbf{ABCD}---see equations~(\ref{eq:xi}) and~(\ref{eq:xi_i})), vertex average degree of $25$ with maximum of $500$, and community sizes varying between $50$ and $1{,}000$. Three different configurations of $(\gamma, \beta)$ guiding the degree distribution and the distribution of cluster sizes are presented on Figure~\ref{fig:timing} ($(\gamma, \beta) \in \{ (3,2), (2.5, 1.5), (2,1)\}$). The number of vertices, $n$, spans from $8{,}192$ to $472{,}392$ and the timings are presented on the log-log scale. We present the results for one run of the {\bf LFR} model whereas averages over five runs of the {\bf ABCD} model are reported. The reason for running the {\bf ABCD} generator more than once was that in most cases one run took less than a second, and so there was some non-negligible variability between runtimes due to external noise when performing the computing.

The conclusion is that the {\bf LFR} algorithm is of the order of 100 times slower than the one for the {\bf ABCD} model; the largest \textbf{ABCD} was generated in a similar time to the smallest \textbf{LFR}. The worst scenario for {\bf ABCD} is when the configuration model is used with low exponents of the two distributions (namely, $\gamma=2$ and $\beta=1$); in this case, {\bf ABCD} is roughly 40 times faster.

\medskip

In order to test an influence of various distributions of $(\gamma, \beta)$ on the of \textbf{ABCD} for larger networks, we performed benchmark tests for $10{,}000{,}000$ vertices. As before, the mixing parameter is fixed to $\mu=0.2$, vertex average degree is $25$ with maximum of $500$, and community sizes vary between $500$ and $10{,}000$ (we increased the community sizes in comparison to the earlier test, as we now consider much larger number of vertices).

The time to generate the graphs using {\bf ABCD} are of order of several minutes---see Table~\ref{tab:timing} where we vary parameters $\gamma$ and $\beta$. In general, Configuration Model variant of {\bf ABCD} is faster when local communities are not very dense. (See the rightmost plot in Figure~\ref{fig:timing} where we presented the case of very dense communities where Chung-Lu based generator is faster.) Also we note that an increase of parameter $\beta$ leads to longer run times. This is associated with the fact that small values of $\beta$ produce several very large communities that attract heavy vertices. In such scenarios, the generators do not have to resolve too many collisions (multiple edges or self loops) and so the algorithm terminates quickly. Each row in Table~\ref{tab:timing} is produced for the same vectors $\mathbf{w}$ and $\mathbf{s}$ (but they vary across rows). The high variability of the results between rows indicates that the run-time is quite sensitive to specific sampled values of $\mathbf{w}$ and $\mathbf{s}$. Specifically, we have checked that the longest run-times are to be expected if there is a lot of heavy vertices sampled in $\mathbf{w}$ and at the same time not many large clusters sampled in $\mathbf{s}$.

\begin{table}[ht]
\begin{center}
\begin{tabular}{|c|r|r|r|r|}
\hline
 $(\gamma,\beta)$ & CL local & CL global & CM local & CM global \\
\hline
 $(3.0,2.0)$ & 170 & 169 & 86 & 94 \\
 $(3.0,1.5)$ & 141 & 184 & 81 & 74 \\
 $(3.0,1.0)$ & 143 & 155 & 85 & 83 \\
 $(2.5,2.0)$ & 228 & 203 & 105 & 118 \\
 $(2.5,1.5)$ & 153 & 132 & 74 & 73 \\
 $(2.5,1.0)$ & 116 & 116 & 67 & 67 \\
 $(2.0,2.0)$ & 167 & 160 & 91 & 91 \\
 $(2.0,1.5)$ & 132 & 132 & 79 & 77 \\
 $(2.0,1.0)$ & 129 & 125 & 72 & 71 \\
\hline
\end{tabular}
\end{center}
\caption{Generation times in seconds of the {\bf ABCD} model---4 variants with $n=10{,}000{,}000$ vertices; CL indicates the Chung-Lu model and CM indicates the configuration model. Generation of comparable graphs with {\bf LFR} would require several hours, which is prohibitive for large-scale studies involving a large number of graphs.
}\label{tab:timing} 
\end{table}

\subsection{Comparing Graph Properties}\label{sec:exp_properties}

In this subsection, we compare graphs generated with the {\bf LFR} and the {\bf ABCD} benchmarks via some topology-based measures. We investigate the following graph statistics: clustering coefficient (the average vertex transitivity), eigenvector centrality, the global transitivity, and the average shortest paths length (approximated via sampling).

\begin{figure}[ht]
\begin{center}
\includegraphics[width=16cm]{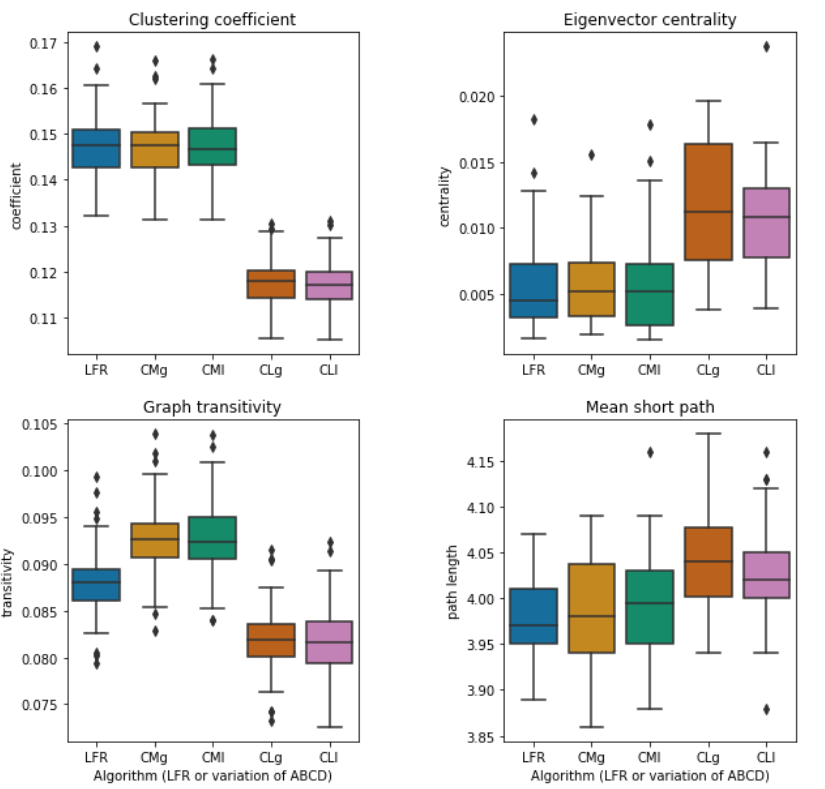}
\end{center}
\caption{Comparing some properties for graphs generated with the {\bf LFR} and {\bf ABCD} benchmarks, using the same degree and community size distributions.
}
\label{fig:properties}
\end{figure}

We generated graphs with 100{,}000 vertices, average degree 25, maximum degree 500 and power law exponent $\gamma = 2.5$; for the community sizes, we used power law exponent $\beta = 1.5$ with sizes between 50 and 2000.
The mixing parameter for {\bf LFR} is set to $\mu=0.2$ and, in order to compare similar graphs,
for the {\bf ABCD} algorithm we derive $\xi$ from~(\ref{eq:xi}) and the $\xi_i$'s from~(\ref{eq:xi_i}) (for the local model).
In Figure \ref{fig:properties}, we report the distribution of the graph properties obtained by generating 30 graphs each using {\bf LFR} as well as 4 variations of {\bf ABCD}, namely:
\begin{itemize}
    \item {\tt CMg}: Configuration Model with global $\xi$,
    \item {\tt CMl}: Configuration Model with local $\xi_i$'s,
    \item {\tt CLg}: Chung-Lu model with global $\xi$,
    \item {\tt CLl}: Chung-Lu model with local $\xi_i$'s.
\end{itemize}

The results of these experiments show high similarity of graphs generated with {\bf LFR} and {\bf ABCD}, in particular, when the configuration model is used. Indeed, some graph parameters that are sensitive with respect to the degree distribution (such as clustering coefficient) are not well preserved for the Chung-Lu variant of the model, which is natural and should be expected. Having said that, all graph parameters we evaluated are relatively well aligned.

\section{Conclusion and Future Work}\label{sec:conclusion}

The paper has two interrelated angles, theoretical and practical. We tried to define the model in as easy and natural way as possible. As a result, from the theoretical point of view, using abundant tools from the theory of random graphs, we plan to investigate an asymptotic behaviour of the \textbf{ABCD} model. As explained in Subsection~\ref{sec:theory_motivation}, this is not only interesting from pure math point of view but also might be important for practitioners. Finally, we plan to generalize the model and add geometry into the model. This would allow, for example, for overlapping and hierarchical communities.

From practical point of view, the implementation we propose in this paper is single-threaded which we believe is sufficient for generating small to medium size graphs. Indeed, it usually takes under one minute to generate a graph consisting of several millions of vertices; in fact, the timing of the process of generating an \textbf{ABCD} graph is of comparable magnitude as the time needed to save it to the hard drive later (on a typical server). However, in order to deal with enormous graphs containing billions of vertices, users might need out-of-core distributed implementation of the \textbf{ABCD} algorithm. In Section~\ref{sec:scalability}, we have commented on how this could be achieved in future work. Independently, it would be interesting to perform more extensive experiments with \textbf{ABCD} (and, in particular, compare it to \textbf{LFR}) when the generated graphs are used to test algorithms that require knowledge of ground truth community structure (such as clustering algorithms). We think that performing such experimental comparison is an important follow-up to this theoretical paper.

\section*{Acknowledgements}
The project is partially financed by the Polish National Agency for Academic Exchange.

\clearpage

\bibliographystyle{plain}
\bibliography{main}

\begin{thebibliography}{10}

\bibitem{DBLP:journals/im/AielloBCJP08}
William Aiello, Anthony Bonato, Colin Cooper, Jeannette C.~M. Janssen, and
  Pawe\l{} Pra\l{}at.
\newblock A spatial web graph model with local influence regions.
\newblock {\em Internet Mathematics}, 5(1):175--196, 2008.

\bibitem{BA}
Reka~Albert Albert-L\'aszl\'o~Barab\'asi.
\newblock Emergence of scaling in random networks.
\newblock {\em Science}, 286(5439):509--512, 1999.

\bibitem{Bae}
Seung-Hee Bae and Bill Howe.
\newblock Gossipmap: A distributed community detection algorithm for
  billion-edge directed graphs.
\newblock {\em In SC’15. ACM}, 27:1–12, 2015.

\bibitem{Barabasi2016}
Albert-L\'aszl\'a Barab\'asi.
\newblock {\em Network Science}.
\newblock Cambridge U Press, 2016.

\bibitem{Bender_Canfield}
Edward~A.\ Bender and E.~Rodney Canfield.
\newblock The asymptotic number of labeled graphs with given degree sequences.
\newblock {\em J.\ Combinatorial Theory Ser.\ A}, 24(3):296–307, 1978.

\bibitem{juliabezanson}
J.~Bezanson, A.~Edelman, S.~Karpinski, and V.B. Shah.
\newblock Julia: A fresh approach to numerical computing.
\newblock {\em SIAM Review}, 69:65--98, 2017.

\bibitem{bollobas2}
B\'{e}la Bollob\'{a}s.
\newblock A probabilistic proof of an asymptotic formula for the number of
  labelled regular graphs.
\newblock {\em European Journal of Combinatorics}, 1:311--316, 1980.

\bibitem{Buzun}
Nazar Buzun, Anton Korshunov, Valeriy Avanesov, Ilya Filonenko, Ilya Kozlov,
  Denis Turdakov, and Hangkyu Kim.
\newblock Egolp: Fast and distributed community detection in billion-node
  social networks.
\newblock {\em IEEE ICDM Mining Workshop}, page 533–540, 2014.

\bibitem{R-MAT}
Deepayan Chakrabarti, Yiping Zhan, and Christos Faloutsos.
\newblock R-mat: A recursive model for graph mining.
\newblock {\em In Proceedings of the 2004 SIAM International Conference on Data
  Mining. SIAM}, page 442–446, 2004.

\bibitem{CL2006}
Fan Chung and Linyuan Lu.
\newblock {\em Complex Graphs and Networks}.
\newblock American Mathematical Society, 2006.

\bibitem{dao_bothorel_lenca}
Vinh~Loc Dao, Cécile Bothorel, and Philippe Lenca.
\newblock Community structure: A comparative evaluation of community detection
  methods.
\newblock {\em Network Science}, page 1–41, 2020.

\bibitem{Emmons}
Scott Emmons, Stephen~G.\ Kobourov, Mike Gallant, and Katy B\"{o}rner.
\newblock Analysis of network clustering algorithms and cluster quality metrics
  at scale.
\newblock {\em PLoS One}, 11:1--18, 2016.

\bibitem{Funke}
Daniel Funke, Sebastian Lamm, Peter Sanders, Christian Schultz, Darren Strash,
  and Mortiz von Looz.
\newblock Communication-free massively distributed graph generation.
\newblock {\em In International Parallel and Distributed Processing Symposium
  (IPDPS)}, 2018.

\bibitem{Paper2}
M.\ Girvan and M.E.J.\ Newman.
\newblock Community structure in social and biological networks.
\newblock {\em Proceedings of the National Academy of Sciences}, 99:7821--7826,
  2002.

\bibitem{Paper19}
Christos Gkantsidis, Milena Mihail, and Ellen~W.\ Zegura.
\newblock The markov chain simulation method for generating connected power law
  random graphs.
\newblock In {\em In ALENEX’03. SIAM}, pages 16--25, 2003.

\bibitem{Greenhill}
Catherine Greenhill and Matteo Sfragara.
\newblock The switch markov chain for sampling irregular graphs and digraphs.
\newblock {\em Theoretical Computer Science}, 719:1–20, 2018.

\bibitem{Hamann:2018:IGM:3178547.3230743}
Michael Hamann, Ulrich Meyer, Manuel Penschuck, Hung Tran, and Dorothea Wagner.
\newblock I/o-efficient generation of massive graphs following the lfr
  benchmark.
\newblock {\em J. Exp. Algorithmics}, 23:2.5:1--2.5:33, August 2018.

\bibitem{Janson}
Svante Janson.
\newblock Random graphs with given vertex degrees and switchings.
\newblock Random Structures Algorithms, to appear, 2019.

\bibitem{Modularity_Pralat}
Bogumi\l{} Kaminski, Valerie Poulin, Pawe\l{} Pra\l{}at, Przemys\l{}aw Szufel,
  and Francois Theberge.
\newblock Clustering via hypergraph modularity.
\newblock {\em PLoS ONE}, 14:e0224307, 2019.

\bibitem{Krioukov}
Dmitri~V.\ Krioukov, Fragkiskos Papadopoulos, Maksim Kitsak, Amin Vahdat, and
  Mari{\'{a}}n Bogu{\~{n}}{\'{a}}.
\newblock Hyperbolic geometry of complex networks.
\newblock {\em Phys.\ Rev.\ E}, 82(036106), 2010.

\bibitem{LFR2}
Andrea Lancichinetti and Santo Fortunato.
\newblock Benchmark graphs for testing community detection algorithms on
  directed and weighted graphs with overlapping communities.
\newblock {\em Physical Review E}, 80, 2009.

\bibitem{LFR}
Andrea Lancichinetti, Santo Fortunato, and Filippo Radicchi.
\newblock Benchmark graphs for testing community detection algorithms.
\newblock {\em Physical Review E}, 78, 2008.

\bibitem{Paper38}
Ron Milo, Nadav Kashtan, Shalev Itzkovitz, Mark~E.J.\ Newman, and Uri Alon.
\newblock On the uniform generation of random graphs with prescribed degree
  sequences.
\newblock {\em arXiv:cond-mat/0312028}, 2003.

\bibitem{Newman_book}
Mark Newman.
\newblock {\em Networks: An Introduction}.
\newblock Oxford University Press, 2010.

\bibitem{Modularity_Newman}
M.E.J.\ Newman and M.\ Girvan.
\newblock Finding and evaluating community structure in networks.
\newblock {\em Phys.\ Rev.\ E.}, 69:26–113, 2004.

\bibitem{Clustering_Pralat}
Liudmila~Ostroumova Prokhorenkova, Pawe\l{} Pra\l{}at, and Andrei Raigorodskii.
\newblock Modularity of complex networks models.
\newblock {\em Internet Mathematics}, 2017.

\bibitem{Paper44}
Jaideep Ray, Ali Pinar, and C.\ Seshadhri.
\newblock Are we there yet? when to stop a markov chain while generating random
  graphs.
\newblock In {\em In WAW’12. Lecture Notes in Computer Science. Springer},
  pages 153--164, 2012.

\bibitem{Paper1}
Fortunato S.
\newblock Community detection in graphs.
\newblock {\em Physics Reports}, 486:75–174, 2010.

\bibitem{Seshadhri2012}
C.\ Seshadhri, Tamara~G.\ Kolda, and Ali Pinar.
\newblock Community structure and scale-free collections of erd\"os-r\'enyi
  graphs.
\newblock {\em Physical Review E}, 85:056109, 2012.

\bibitem{slota_sc2019}
G.~M. Slota, J.~Berry, S.~D. Hammond, S.~Olivier, C.~Phillips, and
  S.~Rajamanickam.
\newblock Scalable generation of graphs for benchmarking hpc
  community-detection algorithms.
\newblock In {\em IEEE International Conference for High Performance Computing,
  Networking, Storage and Analysis ({SC})}, 2019.

\bibitem{Kolda2014}
Kolda T.G., Pinar A., Plantenga T., and Seshadhri C.
\newblock A scalable generative graph model with community structure.
\newblock {\em SIAM Journal on Scientific Computing}, 36:C424--C452, 2014.

\bibitem{vf_cm}
Fabien Viger and Matthieu Latapy.
\newblock Efficient and simple generation of random simple connected graphs
  with prescribed degree sequence.
\newblock In Lusheng Wang, editor, {\em Computing and Combinatorics}, pages
  440--449, Berlin, Heidelberg, 2005. Springer Berlin Heidelberg.

\bibitem{West}
Douglas~B.\ West.
\newblock {\em Introduction to Graph Theory (second edition)}.
\newblock Prentice Hall, 2001.

\bibitem{Winlaw2015}
M.\ Winlaw, H.\ DeSterck, and G.\ Sanders.
\newblock An in-depth analysis of the chung-lu model.
\newblock Technical Report LLNL-TR-678729, Lawrence Livermore Technical Report,
  doi: 10.2172/1239211, 2015.

\bibitem{Wormald}
Nicholas~C.\ Wormald.
\newblock Generating random regular graphs.
\newblock {\em J.\ Algorithms}, 5(2):247–280, 1984.

\bibitem{Zeng}
Jianping Zeng and Hongfeng Yu.
\newblock A study of graph partitioning schemes for parallel graph community
  detection.
\newblock {\em Parallel Computing}, 58:131–139, 2016.

\end{thebibliography}

\clearpage

\section*{Appendix - Algorithm Pseudo-Code}

\begin{figure}[th]
\begin{algorithm}[H]
INPUT: $n$: number of nodes, $\beta$: community sizes power law exponent, $c_{min}$: min community size and $c_{max}$: max community size; $I_{max}$ (optional, default to 100)\;
let $s_{best} := \infty$ and $I := 0$\;
initialize empty list $S_{best}$\;
\Repeat{$I > I_{max}$}
{
check if it is possible to generate the required cluster sizes; throw an error if it is not possible\;
let $s := 0$\;
initialize empty list $X$\;
\Repeat{$s \geq n$}
    {
    Sample value $x$ from truncated discrete power law distribution with parameter $\beta$, restricted to the interval $[c_{min},c_{max}]$ and store in $x$ in $X$\;
    let $s:=s+x$\;
    }

\uIf{$s = n$}{
OUTPUT: list of community sizes $X$\;
{\bf exit};
}
\Else{
\If{$s < s_{best}$}{
let $s_{best}:=s$ and $S_{best} := X$;
}
}
$I=I+1$
}
Truncate $S_{best}$ and update $s_{best}$ accordingly if needed (it might be impossible to find corrections
that produce admissible community sizes in corner cases; this may lead to $s_{best} < n$ case).\;
\Repeat{$s_{best} = n$}{
In random order cyclically precess elements of $S_{best}$\;
If $s_{best} > n$ decrease values sequentially by $1$
unless some element is $c_{min}$; decrease $s_{best}$ by one.\;
If $s_{best} < n$ increase values sequentially by $1$
unless some element is $c_{max}$; increase $s_{best}$ by one.\;
}
OUTPUT: list of community sizes $S_{best}$;
\caption{Generation of the community sizes}
\label{algo:comms}
\end{algorithm}
\end{figure}

\begin{figure}[ht]
\begin{algorithm}[H]
INPUT: $n$: number of nodes, $w_{min}$: min degree, $w_{max}$: max degree, $\gamma$: degree power law exponent; $I_{max}$ (optional, default to 100)\;
initialize empty list $W$\;
let $I := 0$\;
\Repeat{$I > I_{max}$}
{
\Repeat{$|W| = n$}
{
sample value $w$ from truncated discrete power law distribution with parameter $\gamma$, restricted to the interval $[w_{min},w_{max}]$ and add $w$ to $W$\;
}
\If{sum of degrees in $W$ is even}{
OUTPUT: list of degrees $W$
}
let $I := I + 1$
}
decrease the largest value in $W$ by $1$ to make the sum of degrees even\;
OUTPUT: list of degrees $W$
\caption{Generation of the degree sequence}
\label{algo:degrees}
\end{algorithm}
\end{figure}

\begin{figure}[ht]
\begin{algorithm}[H]
INPUT: Degree sequence $W$ on $n$ nodes, community sizes $S$ with $|S|=k$ and parameter $\xi$ (LFR-style $\mu$ can be supplied instead)\;
sort nodes from largest to smallest degrees in W: $w_1 \ge \ldots \ge w_n$\;
sort communities from largest to smallest sizes in S: $s_1 \ge \ldots \ge s_k$\;
initialize number of free spots in each community: $f_i:=s_i,~ 1 \le i \le k$\;
initialize empty lists $S_1, \ldots, S_k$\;
\For{$1 \le i \le n$}{
find max value in $1 \le t \le k$ s.t. $w_i < (1-\xi \phi) s_t$ where $\phi$ is defined in \eqref{eq:phi}\footnote{if $\mu$ is specified instead of $(1-\xi \phi) s_t$, we use $(1-\mu) s_t$.}\;
pick random $1 \le j \le t$ proportional to $f_1,\ldots,f_t$\;
assign vertex $i$ to community $j$ by adding it to $S_j$\;
let $f_j := f_j - 1$\;
}
OUTPUT: community assignment of vertices: $S_1,\ldots,S_k$\;
\caption{Assign nodes with degree sequence $W$ to communities with sizes $S$. Algorithm given for global \textbf{ABCD}. For local version of \textbf{ABCD}, use cluster-local $\xi_i$'s instead of $\xi$.}
\label{algo:assign}
\end{algorithm}
\end{figure}

\clearpage

\begin{figure}[ht]
\begin{algorithm}[H]
INPUT: Community assignment $S_1,\ldots,S_l$ from Algorithm \ref{algo:assign} for $n$ vertices, degree sequence $W$ from  Algorithm \ref{algo:degrees}, and parameter $\xi$ (LFR-style $\mu$ can be supplied instead)\;
if $\mu$ was given, compute $\xi$\;
\For{$1 \le i \le k$}{
let $W_i$, the sum of the degrees of all vertices in $S_i$\;
randomly sample $\lfloor (1-\xi) W_i/2\rceil$ edges within $S_i$ where each vertex is selected proportionally to its internal degree; duplicate edges and self-loops are skipped\;
}
let $s :=$ (sum($W$) - sum($\forall i: W_i$))/2; sample $s$ edges randomly where each vertex is selected proportionally to its external degree; duplicate edges and self-loops are skipped\;
OUTPUT: ABCD graph (list of edges generated)\;
\caption{ABCD with Chung-Lu Model. Algorithm given for global \textbf{ABCD}. For local version of \textbf{ABCD}, use cluster-local $\xi_i$'s instead of $\xi$.}
\label{algo:cm2}
\end{algorithm}
\end{figure}

\vspace{4cm}

\begin{figure}[ht]
\begin{algorithm}[H]
INPUT: Community assignment $S_1,\ldots,S_l$ from Algorithm \ref{algo:assign} for $n$ vertices, degree sequence $W$ from  Algorithm \ref{algo:degrees}, and parameter $\xi$ (LFR-style $\mu$ can be supplied instead)\;
if $\mu$ was given, compute $\xi$\;
\For{$1 \le i \le k$}{
for each vertex in $S_i$, given its degree $w$, assign internal $w_{int} := \lfloor(1-\xi) \cdot w\rceil$\;
if the sum of all $w_{int}$ is odd, adjust highest degree node randomly to make it even\;
wire the vertices in $S_i$ randomly according to their values $w_{int}$\footnote{
other methods can be used here; for example, high degree nodes can be wired first to limit the collisions, or algorithms such as \cite{vf_cm} which yields simple graphs can be used}\;
re-wire duplicated edges and self-loops\;
if re-wiring fails update the $w_{int}$ to achieved values
}
compute the external degree for each vertex as $w_{ext} = :w - w_{int}$\;
wire the vertices randomly according to their values $w_{ext}$ (global model)\;
re-wire duplicated edges and self-loops only considering edges in the global model\;
OUTPUT: ABCD graph (list of edges generated)\;
\caption{ABCD with Configuration Model. Algorithm given for global \textbf{ABCD}.
For local version of \textbf{ABCD}, use cluster-local $\xi_i$'s instead of $\xi$.}
\label{algo:cm1}
\end{algorithm}
\end{figure}

\end{document}